\pdfoutput=1
\documentclass[pdflatex,sn-basic]{sn-jnl}

\usepackage{graphicx}
\usepackage{multirow}
\usepackage{amsmath,amssymb,amsfonts}
\usepackage{amsthm}
\usepackage{booktabs}
\usepackage{tabularx}
\newcolumntype{Y}{>{\raggedright\arraybackslash}X}
\usepackage{xcolor}
\usepackage{tikz}
\usetikzlibrary{arrows.meta,positioning,calc,shapes.geometric}
\usepackage{pgfplots}
\pgfplotsset{compat=1.18}

\definecolor{decblue}{RGB}{31,94,160}
\definecolor{opengray}{RGB}{150,150,150}
\definecolor{okgreen}{RGB}{46,125,50}
\definecolor{badred}{RGB}{183,28,28}
\definecolor{unkgray}{RGB}{224,224,224}

\theoremstyle{thmstyleone}
\newtheorem{proposition}{Proposition}

\usepackage[title]{appendix}
\usepackage[section]{placeins}
\usepackage{needspace}
\hypersetup{bookmarksnumbered=true,bookmarksdepth=2,hypertexnames=false}
\newcommand{\answerbox}[2]{\par\medskip\noindent\fbox{\parbox{\dimexpr\linewidth-2\fboxsep-2\fboxrule\relax}{\textbf{Answer to #1.} #2}}\par\medskip}

\begin{document}

\title[Decided or Open?]{Decided or Open? Comparing Program Repair Techniques When Patch Correctness Is Unknown}

\author*[1]{\fnm{Sadnan} \sur{Nafis}}\email{sadnan.nafis@g.bracu.ac.bd}

\affil*[1]{\orgdiv{Department of Computer Science and Engineering}, \orgname{BRAC University}, \orgaddress{\city{Dhaka}, \country{Bangladesh}}}

\abstract{Comparisons of automated program repair (APR) techniques depend on
knowing which patches are correct, but many patches are never checked and some
archived test results are wrong. We propose reporting each comparison as
decided or open. For two techniques that each pick one patch per bug, we
compute the full range of the difference in their success rates over every
possible value of the unknown labels, and we call the comparison decided when
that range excludes zero. Because a test can only show that a patch is wrong,
more testing can move this range in one direction only. This lets us tell,
before any test runs, which open comparisons testing could never settle. We
study eight settings, including 55,466 candidate patches for Defects4J bugs.
The labels already in the archive decide 12 of 16 comparisons between a
learned patch selector and its baselines, and adding further evidence decides
all 16. In a campaign whose predictions we published in advance, all eight
predictions held. We also show that copies of the developer's fix, which must
pass, are recorded as failing for 33.5\% of GPT-4's copies but for only 2.5\%
of the RepairLLaMA model's, so the reliability of archived test results
differs between models. With these checks, researchers can see which
comparisons need more evidence and which kind of evidence can help.}

\keywords{automated program repair, patch correctness, empirical evaluation,
incomplete labels, partial identification, test oracles}

\maketitle

\section{Introduction}
\label{sec:intro}

Research on automated program repair (APR) is full of comparisons
\citep{legoues2019automated,xia2023automated}. A new model is reported to fix
more Defects4J bugs than the models it was built from
\citep{silva2025repairllama}, and a new patch ranker is said to put a correct
patch first more often than older rankers \citep{bhuiyan2026prevarank}. Each
of these claims counts correct patches, so each depends on knowing which
patches are correct.

In practice, a patch is called correct or incorrect for one of three reasons:
it matches the developer's fix, it fails a test, or a person reviewed it. Many
generated patches meet none of these conditions. A patch that passes every
available test but was never reviewed has no established label, and neither
does a program that compiled in one archived run and failed to compile in an
identical run. Evaluators deal with such patches by convention. They count
them as incorrect, drop them, or trust the tests. These conventions change the
reported results \citep{liu2021critical}. The common ways of labelling patches
disagree with each other \citep{le2019reliability}, and stronger tests alone
have reordered the SWE-bench leaderboard \citep{yu2025utboost,yu2026sweabs}.

\begin{figure}[!htb]
\centering
\begin{tikzpicture}[font=\small, x=0.42cm]
  \draw[-{Latex}] (-6,0) -- (14.5,0) node[right] {difference (pp)};
  \foreach \x in {-5,0,5,10} {
    \draw (\x,0.08) -- (\x,-0.08) node[below] {\ifnum\x>0 $+\x$\else $\x$\fi};
  }
  \draw[dashed, opengray] (0,2.3) -- (0,-0.1);
  \draw[line width=4pt, decblue] (2.84,1.75) -- (11.51,1.75);
  \node[anchor=south, font=\footnotesize, align=center] at (7.2,1.95)
    {selector vs pooled order, Defects4J: \textbf{decided}};
  \draw[line width=4pt, opengray] (-4.41,0.75) -- (9.84,0.75);
  \node[anchor=south, font=\footnotesize, align=center] at (2.7,0.95)
    {selector vs PrevaRank, plausible patches: \textbf{open}};
\end{tikzpicture}
\caption{Two comparisons from our study, shown as decision intervals in
percentage points of success rate. A learned patch selector against the
configurations' pooled generation order on Defects4J is decided by the
archived labels: every possible value of the unknown labels keeps the
difference between $+2.84$ and $+11.51$ points. The same selector against
PrevaRank on plausible patches is open: the difference could be as low as
$-4.41$ or as high as $+9.84$ points, so the unknown labels decide the winner.}
\label{fig:decided-open}
\end{figure}
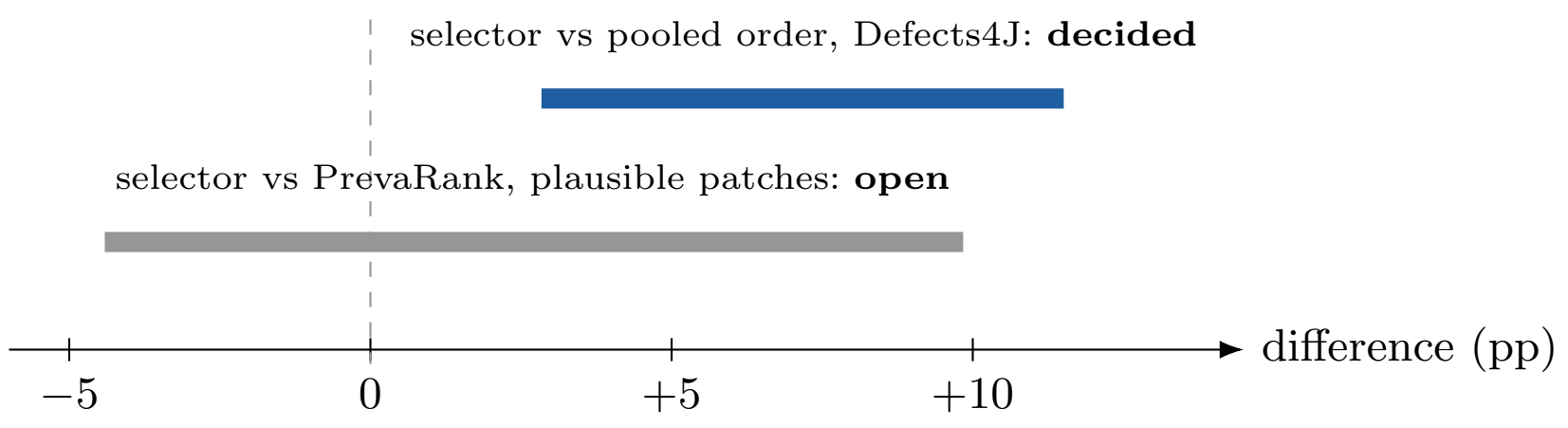

This paper starts from a simple question. Given the labels we have, does the
winner of a comparison depend on the patches whose labels we do not have? For
two techniques that each pick one patch per bug from the same set of
candidates, we compute the smallest and the largest possible difference in
their success rates, trying every possible value for every unknown label. We
call this range the \emph{decision interval}. When it lies entirely on one
side of zero, the comparison is \emph{decided}: no way of filling in the
unknown labels can change the winner. When it contains zero, the comparison is
\emph{open}. Figure~\ref{fig:decided-open} shows one comparison of each kind
from our study. The interval needs no new experiments. It costs one pass over
the labels an archive already holds.

An open comparison needs new evidence, and here the kind of evidence matters.
Most evidence that is cheap to add to an APR archive is testing, and a test
can show that a patch is wrong but can never show that it is right, as
Dijkstra observed for testing in general \citep{dijkstra1972humble}. We show
that this one-sidedness has a sharp consequence. Testing moves the decision
interval toward a single point, the value the comparison takes when every
unknown patch is wrong. If that value is zero, no amount of testing can decide
the comparison (Proposition~\ref{prop:limit}). Because the point can be
computed from the labels already known, we can tell before running anything
which open comparisons a test campaign could ever decide, and whether the
campaign's planned targets include the patches that matter. We call this the
\emph{pre-flight check}.

The check would have saved real effort in our own study. Three test campaigns
on Defects4J used 34.7 container-hours and could not decide one comparison,
because its all-wrong value was zero; a 0.8-hour re-run of the developer's fix
settled it. When we later applied the check before a new campaign and
published its predictions in advance, all eight predictions held.

The test results stored in archives can also be wrong, so we add a check that
needs no execution. Language models often regenerate the developer's fix token
for token, and when the fix lies inside the method the model replaced, such a
copy must pass. Any recorded failure of these copies is therefore an error of
the test harness. In the RepairLLaMA archive, 10.7\% of these copies are
recorded as failing on Defects4J. On the bugs where both models produced
copies, GPT-4's copies are recorded as failing in 33.5\% of cases and the
RepairLLaMA model's in 2.5\%. Other recorded failures of these configurations
are mostly real: when we re-ran a random sample of 160 of them, 147 of the 152
that reached a verdict failed again.

We apply these ideas to eight settings built from five released artifacts and
one new campaign. On RepairLLaMA's Defects4J archive, with 55,466 candidate
patches for 488 bugs, the archived labels already decide 12 of 16 comparisons between a learned patch selector and its baselines. The usual convention, which treats every unlabelled patch as incorrect, would have declared the selector the winner of all 16, by margins of 1.0 to 3.1 points in three of the four comparisons that the archive cannot actually decide. Adding evidence in layers decides all 16, and for twelve of them at least 29 labels would have to be
wrong to reopen the comparison. In most other settings few comparisons are
decided, and the selector that wins every comparison on Defects4J loses to a
simple baseline, occurrence counting, on two other benchmarks.

This paper makes three contributions.
\begin{itemize}
\item \textbf{Decided or open.} A way to report a comparison of APR techniques
as decided or open when many patch labels are unknown, with two measures of
how strong a decision is (Section~\ref{sec:method}).
\item \textbf{The refutation limit and the pre-flight check.} A proof that
evidence which can only show patches wrong cannot decide a comparison whose
all-wrong value is zero, and a check that uses this to predict, before any
test runs, which comparisons a planned campaign can and cannot decide. We
tested the check with predictions published in advance
(Section~\ref{sec:polarity}).
\item \textbf{A free check on the test harness.} Copies of the developer's fix
measure how often an archive records correct patches as failing, without
running anything (Section~\ref{sec:controls}).
\end{itemize}
The study of eight settings is the evidence for these contributions. A
replication package contains the data and scripts behind every number in the
paper.

Section~\ref{sec:background} introduces the problem with a small running
example. Section~\ref{sec:method} presents the method, Section~\ref{sec:design}
the study design, and Section~\ref{sec:results} the results.
Sections~\ref{sec:discussion} to~\ref{sec:conclusion} discuss what the results
mean, relate them to prior work, and examine threats to validity.
\section{Background and a Running Example}
\label{sec:background}

\subsection{How APR Techniques Are Compared}
\label{sec:bg-compare}

An APR technique produces candidate patches for a bug. Modern techniques,
especially those based on language models, produce many candidates per bug,
so they also need a way to choose which candidate to show a developer first.
We call such a rule a \emph{selection policy}. A policy might rank candidates
by how often the same patch was generated, by a learned score, or by the order
in which a model produced them \citep{long2016prophet,bhuiyan2026prevarank}.

The usual measure for comparing two policies is success@1: the share of bugs
whose first chosen patch is correct. Comparing two policies on the same bugs
and the same candidates is a paired comparison. Its result is the difference
between the two success rates, which we report in percentage points.

Computing success@1 requires a correctness label for every chosen patch. Three
sources supply almost all labels in practice:
\begin{itemize}
\item \textbf{Match with the developer's fix.} A candidate whose code is
identical to the developer's fix, or provably equivalent to it, is correct.
\item \textbf{Test failure.} A candidate that fails to compile or fails a test
is incorrect.
\item \textbf{Human review.} A person reads the candidate and judges whether it
is correct.
\end{itemize}

\subsection{Where Labels Go Missing or Wrong}
\label{sec:bg-gaps}

The first two sources are automatic but limited. Matching the developer's fix
only recognises one correct answer, and many correct patches look different.
A test failure shows that a patch is wrong, but a pass shows little, because
tests cover only part of a program's behaviour. Patches that pass all tests
are called \emph{plausible}, and many plausible patches are incorrect
\citep{qi2015plausibility,long2016analysis,smith2015cure}. Human review can
settle these cases, but it is slow, so archives review only some of them.

Two problems follow. First, many patches have \emph{no label}: they pass the
tests and nobody reviewed them. Second, some labels are \emph{wrong}. A
correct patch can be recorded as failing because of a flaky test, a broken
build or a mismatch between the patch and the harness that ran it
\citep{qin2021flaky}. In the archives we study, the same program sometimes
has two archived runs that disagree, one passing and one failing to compile.

Papers usually handle unlabelled patches by convention, for example by
treating every unreviewed plausible patch as incorrect. A single number is then
reported as if all labels were known. Our question is whether the conclusion
of a comparison survives every possible value of the missing labels.

\subsection{A Running Example}
\label{sec:example}

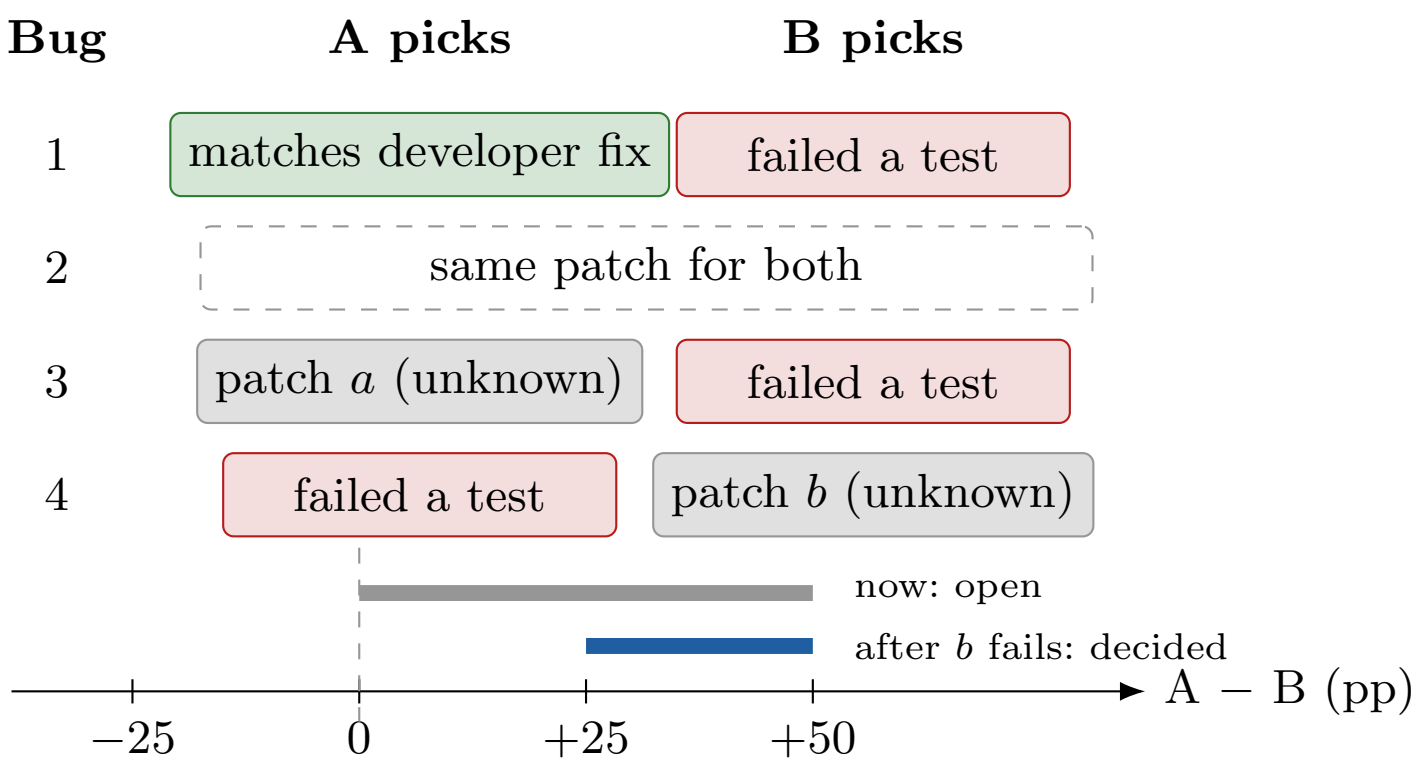
\begin{figure}[t]
\centering
\begin{tikzpicture}[
  font=\small,
  pt/.style={draw, rounded corners=2pt, minimum width=2.6cm, minimum height=0.55cm, align=center},
  ok/.style={pt, fill=okgreen!20, draw=okgreen},
  bad/.style={pt, fill=badred!15, draw=badred},
  unk/.style={pt, fill=unkgray, draw=opengray},
  same/.style={pt, fill=white, draw=opengray, dashed}]
  \node[font=\small\bfseries] at (0,0) {Bug};
  \node[font=\small\bfseries] at (2.4,0) {A picks};
  \node[font=\small\bfseries] at (5.4,0) {B picks};
  \node at (0,-0.75) {1};
  \node[ok]  at (2.4,-0.75) {matches developer fix};
  \node[bad] at (5.4,-0.75) {failed a test};
  \node at (0,-1.5) {2};
  \node[same] at (3.9,-1.5) {\hspace{1.4cm}same patch for both\hspace{1.4cm}};
  \node at (0,-2.25) {3};
  \node[unk] at (2.4,-2.25) {patch $a$ (unknown)};
  \node[bad] at (5.4,-2.25) {failed a test};
  \node at (0,-3.0) {4};
  \node[bad] at (2.4,-3.0) {failed a test};
  \node[unk] at (5.4,-3.0) {patch $b$ (unknown)};
  \begin{scope}[yshift=-4.3cm]
    \draw[-{Latex}] (-0.3,0) -- (7.2,0) node[right] {A $-$ B (pp)};
    \foreach \x/\l in {0.5/$-25$,2.0/0,3.5/$+25$,5.0/$+50$} {
      \draw (\x,0.08) -- (\x,-0.08) node[below] {\l};
    }
    \draw[dashed, opengray] (2.0,0.95) -- (2.0,-0.2);
    \draw[line width=3pt, opengray] (2.0,0.65) -- (5.0,0.65);
    \node[anchor=west, font=\footnotesize] at (5.15,0.65) {now: open};
    \draw[line width=3pt, decblue] (3.5,0.3) -- (5.0,0.3);
    \node[anchor=west, font=\footnotesize] at (5.15,0.3) {after $b$ fails: decided};
  \end{scope}
\end{tikzpicture}
\caption{The running example. Two policies pick one patch for each of four
bugs. Green patches are known to be correct, red patches are known to be
wrong, and grey patches $a$ and $b$ passed all tests but were never reviewed.
Below, the range of the difference in success rate (A minus B) before and
after a test shows that $b$ is wrong.}
\label{fig:example}
\end{figure}

Figure~\ref{fig:example} shows two policies, A and B, on four bugs. Each
policy picks one patch per bug from the same candidates.

\begin{itemize}
\item On bug 1, A picks a patch that matches the developer's fix, and B picks
a patch that failed a test. A wins this bug.
\item On bug 2, both policies pick the same patch. Whatever its label, it
cannot separate them.
\item On bug 3, A picks patch $a$, which passed all tests but was never
reviewed, and B picks a patch that failed.
\item On bug 4, A picks a patch that failed, and B picks patch $b$, which
also passed all tests without review.
\end{itemize}

Write $a=1$ if patch $a$ is correct and $a=0$ if it is wrong, and the same for
$b$. A scores on bug 1 and possibly on bug 3; B scores possibly on bug 4.
Bug 2 adds the same amount to both. The difference in success rate is
therefore
\[
  \text{A} - \text{B} \;=\; \frac{1 + a - b}{4}.
\]
The difference is smallest when $a$ is wrong and $b$ is correct, which gives
0, and largest when $a$ is correct and $b$ is wrong, which gives $2/4$. The
decision interval is therefore $[0, +50]$ percentage points. It touches zero,
so the comparison is \emph{open}: if $a$ is wrong and $b$ is correct, the two
policies tie, and the archive cannot tell us which one is better.

\paragraph{Which test helps} Suppose we can run one more test on one of the
two unknown patches. If the test shows that $b$ is wrong, the difference
becomes $(1+a)/4$, which lies between $+25$ and $+50$ points whatever $a$ is.
The comparison is now decided in favour of A. If instead the test shows that
$a$ is wrong, the difference becomes $(1-b)/4$, which still ranges from 0 to
$+25$ points, and the comparison stays open. Only one of the two patches is
worth testing for this comparison, and which one it is follows from the labels
we already have.

\paragraph{When no test can help} Now remove bug 1 from the example. The
difference becomes $(a-b)/3$. If every unknown patch turns out to be wrong,
the difference is exactly zero. A test can only set an unknown patch to wrong,
so after any number of tests the best possible outcome is still a tie, and the
comparison can never be decided by testing. It can be decided only by evidence
that a patch is \emph{correct}, for example a review showing that $a$ is
correct, together with a test showing that $b$ is wrong.

This small example contains the three ideas of the paper. The decision
interval tells us whether the winner depends on unknown labels. Its value when
every unknown patch is wrong, which is $+25$ points with bug 1 and zero
without it, tells us whether testing can ever settle the comparison. The
labels we already know tell us which patches a test campaign should target.
Section~\ref{sec:method} states these ideas in general.
\section{Deciding a Comparison When Labels Are Missing}
\label{sec:method}

This section turns the running example into a general method. We first define
the decision interval (Section~\ref{sec:intervals}) and the rules for
computing it on a real archive (Section~\ref{sec:contract}). We then measure
how strong a decision is (Section~\ref{sec:strength}), show what testing can
and cannot decide (Section~\ref{sec:polarity}), and describe a check on the
test harness that needs no execution (Section~\ref{sec:controls}).

\subsection{The Decision Interval}
\label{sec:intervals}

A setting contains $N$ bugs. For each bug $j$ there is a pool of generated
candidate patches. Candidates with the same abstract syntax tree are the same
program, so we group them into one class and give the whole class one label.
Candidates that do not parse are grouped by exact text. We write
$\mathcal{C}_j$ for the classes of bug $j$.

A selection policy scores every class without seeing any correctness evidence
and proposes its highest-scoring class. When several classes share the top
score, it picks one of them uniformly at random. Its behaviour on bug $j$ is
therefore a probability for each class: policy $A$ proposes class $c$ with
probability $\alpha_c$, and policy $B$ with probability $\beta_c$. A policy
with no eligible class abstains and scores zero on that bug.

Let $y_c=1$ if class $c$ is correct and $y_c=0$ if it is not. With every bug
weighted equally, the difference in success@1 between $A$ and $B$ is
\begin{equation}
  \Delta \;=\; \sum_{j=1}^{N}\frac{1}{N}\sum_{c\in\mathcal{C}_j}(\alpha_c-\beta_c)\,y_c
        \;=\; \sum_c d_c\,y_c,\qquad d_c=\frac{\alpha_c-\beta_c}{N}.
  \label{eq:delta}
\end{equation}
The weight $d_c$ says how much class $c$ matters to the comparison. It is
positive when $A$ proposes the class more often than $B$, negative in the
opposite case, and zero when both propose it equally often. A class with
$d_c=0$ cannot affect the comparison, whatever its label, like the shared
patch of bug 2 in the running example.

Let $K$ be the classes whose label is known and $U$ the classes whose label is
unknown. Because $\Delta$ is a weighted sum of labels, its smallest value sets
every unknown label against $A$ and its largest value sets every unknown label
in favour of $A$:
\begin{equation}
  L=\sum_{c\in K}d_c y_c+\sum_{c\in U}\min(0,d_c),\qquad
  H=\sum_{c\in K}d_c y_c+\sum_{c\in U}\max(0,d_c).
  \label{eq:bounds}
\end{equation}
Both values are reached by actual assignments of the unknown labels, so the
interval $[L,H]$ is exactly the set of differences the evidence allows when
nothing is assumed about the unknown patches. The comparison is
\emph{decided} when the interval excludes zero ($L>0$ or $H<0$), and
\emph{open} otherwise. In the language of partial identification, a decided
comparison is one whose sign is identified \citep{kasy2016partial}. We report
$L$ and $H$ in percentage points (pp) and compute them with exact rational
arithmetic.

The interval is Manski's bound without assumptions \citep{manski2003partial},
applied to the difference $\Delta$. What is new here is making it work on APR
archives. This requires grouping identical patches, pooling a bug's candidates
across prompts, handling conflicting archived results, and separating evidence
that can confirm a patch from evidence that can only refute it. Because both
policies share the same labels, the interval is tighter than the difference of
two separate ranges, one per policy, such as the residual of rank-biased
precision \citep{moffat2008rbp}.

Filling in every unknown label the same way does not give a bound. Suppose two
policies propose different unreviewed patches $p$ and $q$. Labelling both
incorrect, or both correct, makes the policies tie, yet the true difference
could be $+1$ (only $p$ is correct) or $-1$ (only $q$ is correct).

Two particular assignments play a central role in what follows. The
\emph{all-wrong value} sets every unknown label to incorrect, and the
\emph{all-correct value} sets every unknown label to correct:
\begin{equation}
  \Delta_0=\sum_{c\in K}d_c y_c,\qquad \Delta_1=\Delta_0+\sum_{c\in U}d_c .
  \label{eq:a0a1}
\end{equation}
Both always lie inside $[L,H]$. In the running example, $\Delta_0$ is $+25$
points with bug 1 and zero without it.

\subsection{Rules for Applying the Interval to an Archive}
\label{sec:contract}

The interval depends on a few definitions: what counts as one bug, when two
patches are the same, how conflicting labels are handled, and how ties are
broken. We fix these rules once and apply them to every setting except three
that keep the units of their original analyses (Section~\ref{sec:settings}).
Section~\ref{sec:rq1} varies the rules that could matter.

\paragraph{One bug, one decision} The unit is the bug. Some archives show the
same bug to a model through several prompts; RepairLLaMA presents 379 of its
488 Defects4J bugs in two formats. We pool the candidates of all prompts,
because a developer facing the bug faces one decision. Bugs with no candidates
stay in the denominator and count as zero for both policies.

\paragraph{Identical patches share a label} Candidates are grouped by a
fingerprint of their Java syntax tree with comments and whitespace removed. A
class takes the label its labelled members agree on. Sometimes they disagree,
for example when one copy of a program was archived as passing and an
identical copy as failing to compile. We call such a class
\emph{conflict-locked}. It counts as unknown, and adding one more test result
to one of its members cannot unlock it. Evidence about the program as a whole,
such as a human verdict, can unlock it, and so can showing that the
conflicting results were wrong.

\paragraph{Unchanged code} A candidate identical to the buggy method changes
nothing. Every buggy version we executed failed its bug-revealing tests, so an
unlabelled candidate of this kind is labelled incorrect. The rule applies to
both policies in the same way.

\paragraph{Ties} Policies see every non-empty candidate, whether or not it
parses, and break ties uniformly at random.

\subsection{How Strong Is a Decision?}
\label{sec:strength}

A decided comparison relies on the labels we already know, and some of those
labels could be wrong. Two measures tell a reader how much weight the decision
can bear. We state them for $L>0$; the case $H<0$ is symmetric.

\paragraph{Flips to reopen} Changing a known label $y_c$ to its opposite
changes $\Delta$ by $d_c(1-2y_c)$. The harmful changes are those that turn a
correct class with $d_c>0$ into an incorrect one, or an incorrect class with
$d_c<0$ into a correct one. Each lowers $L$ by $|d_c|$. Taking the harmful
changes from the largest $|d_c|$ downward and stopping when their sum reaches
$L$ gives the exact minimum number of wrong labels that would reopen the
comparison. A decision that needs 50 wrong labels to reopen is robust; one
that needs 2 is fragile.

\paragraph{The labels a decision rests on} Some decided comparisons rest on
a few labels from a source that is easy to doubt, such as human judgment. To
show this, we find a smallest set of known labels that is enough on its own
to keep the decision, given that every other label is treated as unknown. We
call it the \emph{decision basis}. Formally, write $s\in\{+1,-1\}$ for the
sign of a decided comparison, and let $M$ be $L$ when $s=+1$ and $-H$ when
$s=-1$, so that the decision holds exactly when $M>0$. With every label
unknown, $M$ equals $\sum_c\min(0,s\,d_c)$, and learning the label of class
$c$ adds $g_c=s\,d_c y_c-\min(0,s\,d_c)\ge 0$. A set $S$ of known labels is a
decision basis when $\sum_c\min(0,s\,d_c)+\sum_{c\in S}g_c>0$. We order the
label sources from most to least objective:
\begin{enumerate}
\item match with the developer's fix;
\item execution failure, archived or from our controlled re-execution;
\item labels set by the re-run of the developer's fix (Section~\ref{sec:layers});
\item human judgment;
\item differences found by generated tests;
\item the unchanged-code rule.
\end{enumerate}
We then build the basis that uses as few labels as possible from the last
source in this order, then from the one before it, and so on. This shows, for
example, whether a decision needs any human verdict at all. We also report the
\emph{break-even error rate} $M/\sum_c g_c$. If every supporting label were
wrong independently with this probability, the expected margin would be zero.

\subsection{What Testing Can and Cannot Decide}
\label{sec:polarity}

Evidence sources differ in which labels they can set. A test run can show that
a patch is wrong, but a passing run leaves its correctness unknown. Generated
tests act in the same direction: a test on which a patch and the developer's
fix behave differently can mark the patch incorrect, and never correct. Human
review and proven equivalence to the developer's fix can establish that a
patch is correct. We call evidence \emph{refuting} if it can only change
unknown labels to incorrect, and \emph{confirming} if it can only change them
to correct. A third kind, \emph{retracting} evidence, removes a known label,
for instance when a re-run shows that an archived failure does not reproduce.
Retraction on its own only widens an interval. It decides a comparison only
when it unlocks a conflict-locked class whose remaining labels are correct,
and it then acts as confirmation.

The following result makes the running example general.

\begin{proposition}[Refutation limit]
\label{prop:limit}
Let $R\subseteq U$ be the unknown classes that an evidence source can refute,
and suppose the source changes labels only by setting classes in $R$ to
incorrect. Then $\Delta_0$ does not change, $L$ never decreases, $H$ never
increases, and $L\le \Delta_0\le H$ throughout. The source can decide the
comparison in favour of $A$ if and only if
$\Delta_0+\sum_{c\in U\setminus R}\min(0,d_c)>0$, and in favour of $B$ if and
only if $\Delta_0+\sum_{c\in U\setminus R}\max(0,d_c)<0$. In particular, it
can never decide a comparison with $\Delta_0=0$. In the first case, the
fewest refutations that decide the comparison are the classes $c\in R$ with
$d_c<0$, taken from the largest $|d_c|$ downward until their total exceeds
$\sum_{c\in U,\,d_c<0}|d_c|-\Delta_0$. The second case is symmetric.
\end{proposition}

\begin{proof}
$\Delta_0$ sums $d_c y_c$ over known classes only. A refuted class joins $K$
with $y_c=0$ and adds nothing, so $\Delta_0$ is unchanged. Its term in $L$
moves from $\min(0,d_c)$ to $0$, and its term in $H$ from $\max(0,d_c)$ to
$0$. Once every class in $R$ is refuted,
$L=\Delta_0+\sum_{c\in U\setminus R}\min(0,d_c)$ and
$H=\Delta_0+\sum_{c\in U\setminus R}\max(0,d_c)$, which are the best values
the source can reach. Each refutation moves the bounds independently of the
others, so taking the largest $|d_c|$ first minimises their number.
\end{proof}

In words, refuting evidence can only pull the interval toward the all-wrong
value $\Delta_0$. Since $\Delta_0$ is the net weight of the patches already
known to be correct, it is zero whenever both policies propose every
known-correct patch equally often. A common case is two policies that both
pick the patch matching the developer's fix whenever the pool contains one.
No amount of testing can separate such policies. Confirming evidence obeys the
mirror statement, with the all-correct value $\Delta_1$ in place of
$\Delta_0$.

Two practical conditions apply. First, a refutation that lands in a class with
a known-correct member does not set a new label; it creates a conflict and
removes a known label. Execution evidence is therefore refuting only when no
refutation falls in such a class, and we check this for every campaign.
Second, the set $R$ depends on the campaign as well as on the kind of
evidence. A conflict-locked class cannot be refuted by labelling one more of
its members. Our Defects4J re-runs therefore left to new tests and human
judgment the classes whose members had all passed the archived tests, since a
re-run could still fail them through flakiness, a wrong archived record or a
changed environment.

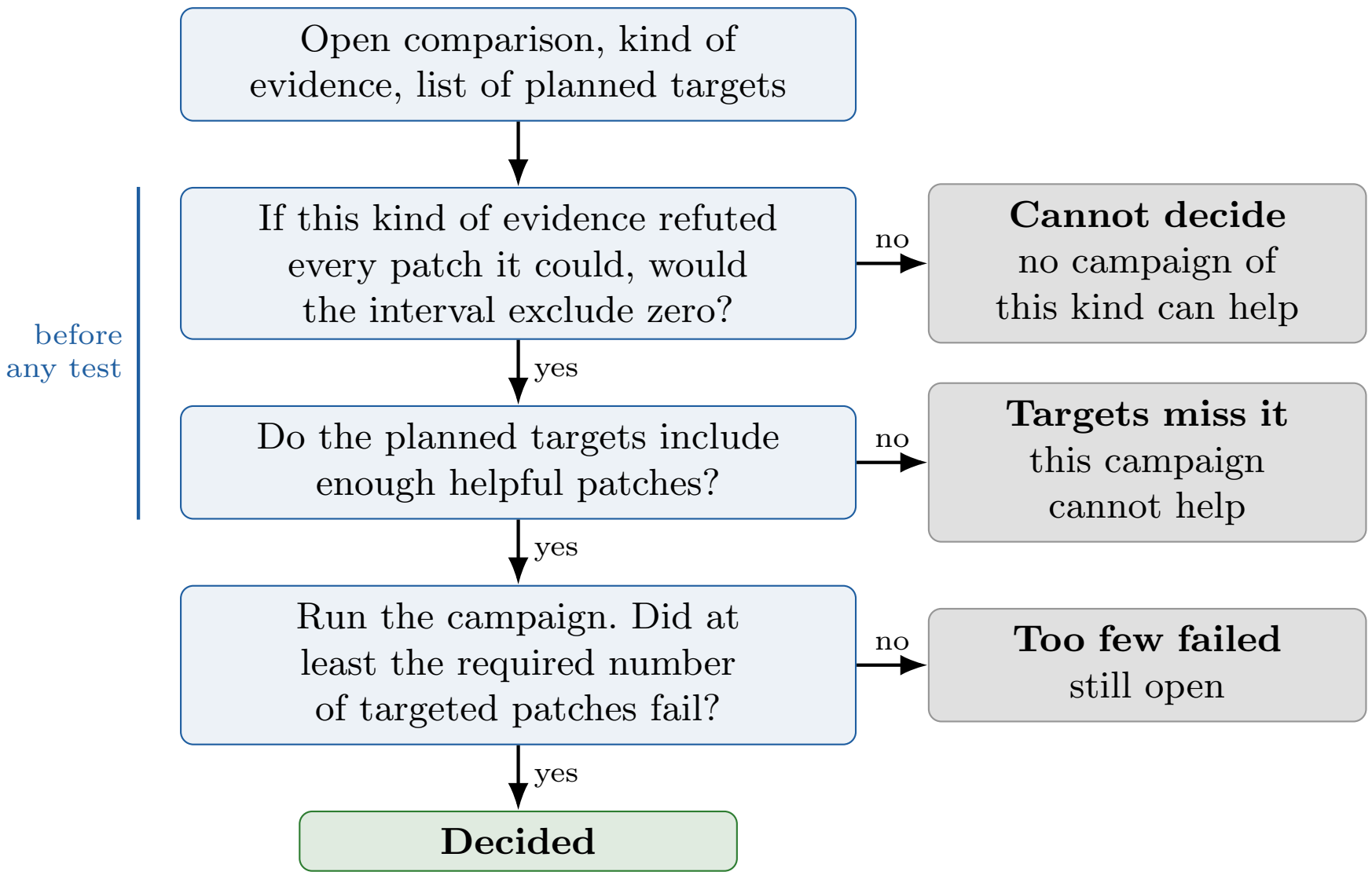
\begin{figure}[t]
\centering
\providecommand\nohyph{\hyphenpenalty=10000\relax}
\begin{tikzpicture}[
  font=\small,
  every node/.append style={execute at begin node=\nohyph},
  node distance=0.55cm and 0.6cm,
  q/.style={draw, rounded corners=3pt, fill=decblue!8, draw=decblue, text width=5.4cm, align=center, inner sep=4pt},
  endbox/.style={draw, rounded corners=3pt, fill=unkgray, draw=opengray, text width=3.4cm, align=center, inner sep=4pt},
  win/.style={draw, rounded corners=3pt, fill=okgreen!15, draw=okgreen, text width=3.4cm, align=center, inner sep=4pt},
  arr/.style={-{Latex}, thick}]
  \node[q] (start) {Open comparison, kind of evidence, list of planned targets};
  \node[q, below=of start] (q1) {If this kind of evidence refuted every patch it could, would the interval exclude zero?};
  \node[endbox, right=of q1] (o1) {\textbf{Cannot decide}\\ no campaign of this kind can help};
  \node[q, below=of q1] (q2) {Do the planned targets include enough helpful patches?};
  \node[endbox, right=of q2] (o2) {\textbf{Targets miss it}\\ this campaign cannot help};
  \node[q, below=of q2] (q3) {Run the campaign. Did at least the required number of targeted patches fail?};
  \node[endbox, right=of q3] (o3) {\textbf{Too few failed}\\ still open};
  \node[win, below=of q3] (o4) {\textbf{Decided}};
  \draw[arr] (start) -- (q1);
  \draw[arr] (q1) -- node[above, font=\footnotesize] {no} (o1);
  \draw[arr] (q1) -- node[right, font=\footnotesize] {yes} (q2);
  \draw[arr] (q2) -- node[above, font=\footnotesize] {no} (o2);
  \draw[arr] (q2) -- node[right, font=\footnotesize] {yes} (q3);
  \draw[arr] (q3) -- node[above, font=\footnotesize] {no} (o3);
  \draw[arr] (q3) -- node[right, font=\footnotesize] {yes} (o4);
  \draw[decblue, thick] ([xshift=-0.35cm]q1.north west) -- ([xshift=-0.35cm]q2.south west)
    node[midway, left, align=right, font=\footnotesize, text=decblue] {before\\ any test};
\end{tikzpicture}
\caption{The pre-flight check for one open comparison and one planned test
campaign. The first two questions are answered before anything runs. The last
one is answered by running the campaign; its chance of success can be
estimated beforehand from the yield of a small pilot.}
\label{fig:preflight}
\end{figure}

\paragraph{The pre-flight check} The proposition gives a check that runs
before any test, using only the current labels, the policies' choices and the
list of patches a campaign plans to run. Figure~\ref{fig:preflight} shows it
as a flow chart. The check first asks whether evidence of this kind could
decide the comparison even if it refuted every patch it is able to refute. If
not, no campaign of this kind can help. If it could, the check asks whether
the campaign's planned targets include enough \emph{helpful} patches, those
whose refutation moves the bound that has to cross zero. If they do not, this
particular campaign is wasted on this comparison. If they do, the proposition
gives the smallest number of targeted patches that must fail.

Whether enough targeted patches then fail is known only after the campaign.
Its chance can be estimated in advance. If each targeted helpful patch fails
independently with a rate measured on a small pilot share of the targets, or
taken from an earlier campaign, simulating the refuted weight gives the
probability that the campaign decides the comparison. We call this simple
simulation the \emph{planner}.

The impossibility half of the proposition holds even if some refutations are
wrong, because $\Delta_0$ depends only on labels that were already known. The
other half assumes the refutations are right. A decision reached by refutation
survives any $m$ false refutations exactly when the refuted weight exceeds
what the comparison needs by at least the $m$ largest refuted weights.

\subsection{A Free Check on the Test Harness}
\label{sec:controls}

Test results can be wrong, and a wrong failure removes a correct patch from
the comparison. Benchmarks guard against this by running the developer's fix
and checking that it passes \citep{jimenez2024swebench}. Released repair
archives rarely contain that run, but they contain a substitute. Language
models often regenerate the developer's fix exactly. When the developer's fix
lies entirely inside the method that a candidate replaces, such a copy must
compile and pass, provided the baseline is correct, the repair context is
complete and the tests are fair. The share of these copies recorded as failing
therefore measures how often the harness rejects correct patches, without
running anything.

We identify copies by comparing their tokens with the developer's fix after
removing comments and whitespace. We keep only bugs whose official fix touches
nothing outside the replaced method. To compare two models fairly, we use
only the bugs where both produced copies, average each bug's failure rate with
equal weight per bug, and resample whole bugs in a paired bootstrap, so that
repeated copies for one bug do not count as independent evidence.
Re-executing the copies recorded as failing shows how many of those failures
actually reproduce (Section~\ref{sec:rq3}).
\section{Study Design}
\label{sec:design}

We ask three research questions.
\begin{description}
\item[RQ1] How often are APR comparisons decided by the evidence we have, and
what does each kind of added evidence contribute?
\item[RQ2] Can we predict, before running tests, which comparisons a test
campaign will settle?
\item[RQ3] How trustworthy are the archived test results that the decisions
rest on?
\end{description}

\subsection{Settings}
\label{sec:settings}

Table~\ref{tab:settings} lists the eight settings. The main one is the
Defects4J evaluation \citep{just2014defects4j} released with RepairLLaMA
\citep{silva2025repairllama}. Fourteen model configurations produced up to ten
patches for each of 488 single-function bugs. The configurations are base,
fine-tuned and adapter versions of CodeLlama-7B and DeepSeek-Coder-6.7B, plus
GPT-3.5 and GPT-4. Together they produced 64,992 non-empty outputs, which
contain 55,466 distinct candidates. In 483 of the 484 bugs that have
candidates, the developer's fix changes only the method that the candidates
replace, so a candidate can in principle be correct on its own.

The archive labels 1,544 candidates correct: 1,305 because they match the
developer's fix and 239 by human review. It labels 51,751 incorrect: 50,449
because of a recorded compile or test failure and 1,302 by human review. The
remaining 2,171 candidates, 3.9\% of the total, have no label. Every human
label in the archive comes from a completed review by two agreeing reviewers,
or with a third reviewer breaking a tie.

\begin{table}[t]
\caption{The eight settings. A comparison is between two selection policies
on one set of candidates.}
\label{tab:settings}
\small
\begin{tabularx}{\linewidth}{@{}lYrrr@{}}
\toprule
Setting & Where the candidates come from & Candidates & Bugs & Comparisons \\
\midrule
Defects4J & RepairLLaMA archive, 14 model configurations & 55,466 & 488 & 16 \\
HumanEval-Java & RepairLLaMA archive, 14 model configurations & 16,925 & 162 & 6 \\
Fresh campaign & our own run of Qwen2.5-Coder-7B-Instruct & 2,628 & 488 & 5 \\
RepairBench & GitBug-Java outputs of 35 models & 21,144 & 89 & 9 \\
D4C & outputs released with D4C & 2,594 & 276 & 4 \\
Plausible patches & Defects4J candidates that pass all tests & 2,130 & 488 & 21 \\
PrevaRank input order & plausible RepairLLaMA candidates & 4,180 & 227 & 51 \\
Overfitting detectors & benchmark of \citet{williams2026pod} & 169 & 50 & 10 \\
\bottomrule
\end{tabularx}
\end{table}

The same archive contains RepairLLaMA's evaluation on HumanEval-Java
\citep{jiang2023impact}, a benchmark of small single-method tasks. The other
settings test whether the picture holds elsewhere. They are RepairBench's
leaderboard outputs on GitBug-Java \citep{silva2025repairbench,silva2024gitbugjava},
the candidates released with D4C \citep{xu2024d4c}, the plausible Defects4J
patches that remain once tests have run, the input orders given to PrevaRank
\citep{bhuiyan2026prevarank}, and the overfitting-detector benchmark of
\citet{williams2026pod}.

The fresh campaign is our own. We ran Qwen2.5-Coder-7B-Instruct
\citep{hui2024qwen25coder}, a model that is not among RepairLLaMA's
configurations, to generate ten patches for each Defects4J bug whose buggy
method the archive holds, showing it only the buggy method. Before drawing the first sample, we fixed in writing
which policies we would compare, which comparisons we would report, how
evidence would be collected, and when we would stop.

\subsection{Policies}
\label{sec:policies}

The \emph{learned selector} was trained on the archive's metadata before this
study and has not changed since. It estimates three probabilities for each
candidate: that it compiles, that it passes its tests if it compiles, and that
it is correct if it passes. It ranks candidates by the product of the three.
Each estimate comes from a logistic regression over 46 features that describe
where a candidate appears: its positions in the configurations' lists of ten
samples, how many configurations produced it, and which ones. The features
contain no code and no labels. The models are fitted on bug-disjoint folds,
so every candidate is scored by a model that never saw its bug. For the other
settings, we refit the selector once on all 488 Defects4J bugs and apply it
unchanged.

The selector is the vehicle of this study, not its subject. Unknown labels matter most in comparisons between close competitors, and a selector with several ablations and many baselines supplies 16 such comparisons on one pool. On Defects4J we compare the selector with 16 baselines in four groups:
\begin{itemize}
\item \textbf{Order-based.} Six baselines combine the configurations' orders.
The main one is the \emph{pooled order}, which proposes a candidate that some
configuration ranked first (minimum-rank aggregation, MRA). The others are the
first position in the concatenated lists, Borda counts, reciprocal-rank
fusion, mean normalised position, and the top candidate of the single best
configuration.
\item \textbf{Agreement-based.} \emph{Occurrence count} proposes the patch that
most configurations generated, a form of self-consistency voting
\citep{wang2023selfconsistency}. A second baseline uses the pooled order with
occurrence count as a tie-breaker.
\item \textbf{Ablations of the selector.} The product of its first two
estimates (the \emph{testability product}), a single-stage model, and a refit
without configuration identities (the \emph{source-agnostic refit}).
\item \textbf{Code-reading.} Token similarity and CodeT5+ embedding similarity
to the buggy method \citep{wang2023codet5plus}, and naturalness and entropy
change under a 1.5B code model \citep{yang2024naturalness}.
\end{itemize}
The last baseline is uniform random choice. For plausible patches we add the
selector's correctness stage, APPT \citep{zhang2024appt}, which predicts
whether a plausible patch overfits, and PrevaRank \citep{bhuiyan2026prevarank}.
The other settings use the policies their data support.

\subsection{Evidence Added on Defects4J}
\label{sec:layers}

We added evidence in five layers and recomputed every comparison after each
one. Table~\ref{tab:layers} lists them. The column ``can show patches'' gives
the direction of each layer in the sense of Section~\ref{sec:polarity}.

\begin{table}[t]
\caption{Evidence layers on Defects4J, in the order applied. ``Can show
patches'' is the direction in which a layer can change labels. Hours are
container-hours of execution. ``Decided'' counts the decided comparisons
after the layer, out of 16.}
\label{tab:layers}
\small
\begin{tabularx}{\linewidth}{@{}lYlrr@{}}
\toprule
Layer & What it adds & Can show patches & Hours & Decided \\
\midrule
E0 Archive & labels already in the archive & correct or wrong & -- & 12 \\
E1 Census & re-run of 280 candidates with conflicting records & wrong & 16.0 & 13 \\
E2 Top-up & re-run of 179 further relevant groups & wrong & 5.7 & 13 \\
E3 Generated tests & EvoSuite and Randoop tests for 165 test-passing groups & wrong & 13.0 & 14 \\
E4 Review & two people judge 44 test-passing groups & correct or wrong & -- & 15 \\
E5 Fix re-run & re-run of 80 archived failures of copies of the developer's fix & correct & 0.8 & 16 \\
\bottomrule
\end{tabularx}
\end{table}

\paragraph{Re-running candidates (E1, E2, E5)} We re-ran candidates in a
pinned container with Defects4J 3.0.1 and Java 11. Each run executes the
developer's fixed version, the buggy version and the candidate on the bug's
failure-revealing tests, twice, the second time in reverse order. A test
counts only if the fixed version passes it twice and the buggy version fails
it twice in the same way. A candidate that fails such a test twice in the same
way is marked wrong. A candidate that passes stays unknown, because passing
does not show that a patch is correct.

\paragraph{Generated tests (E3)} For 165 relevant groups whose members all
passed the archived tests, EvoSuite \citep{fraser2011evosuite} and Randoop
\citep{pacheco2007feedback} generated tests on the developer's fixed version.
A generated test that passes on the fix and fails on the candidate in two runs
marks the candidate wrong, unless the difference falls in one of four cases
fixed in advance: the environment, the text of an exception message, output
that the specification leaves open (such as hash codes or the order of an
unordered collection), or mocked time, randomness, files or network.

\paragraph{Human review (E4)} Two annotators, the author and a second
annotator, judged the 44 relevant groups that were still unknown after the
census and had passed all archived tests. These were mixed with 60 archived
human judgments as an audit. The annotators saw the buggy method, the
developer's fix and the candidate. They used a deliberately strict rule: a
candidate counts as correct only if it behaves like the fix on every input
that can reach the method. They did not see which comparison a candidate
mattered for, which items were audit items, or the archived labels, and each
annotator labelled without seeing the other's labels. The two agreed on all
104 items.

\paragraph{Re-running copies of the developer's fix (E5)} Some candidates
are copies of the developer's fix that differ only in local names or
annotations, so exact matching misses them. We found 40 groups that contain
such a copy but carry archived failures, and we re-ran those 80 failures. A
failure that does not reproduce is withdrawn. A group that then has no
remaining label and contains a copy of the fix becomes correct.

\paragraph{Use of AI tools} An AI assistant helped write and run the analysis scripts, and it screened the counterexamples of the generated-test campaigns against the four exclusions fixed in advance. Every screening verdict and its reason is recorded in the replication package, and the author checked and is accountable for all analyses and text.

Appendix~\ref{app:procedures} gives the full procedures, including those for
HumanEval-Java, the fresh campaign and D4C.

\subsection{Checks on the Evidence Itself}
\label{sec:checks}

Several checks measure the evidence rather than add to it:
\begin{itemize}
\item We re-ran the census candidates in the original Defects4J 2.0 and Java 8
environment.
\item We ran 100 randomly chosen candidates that the archive accepts as
correct through the same protocol, looking for false failures.
\item We counted the archive's copies of the developer's fix that are recorded
as failing (Section~\ref{sec:controls}).
\item A \emph{failure audit} re-ran a random sample of 160 archived failures
from the five configurations whose fix copies fail more than 10\% of the time
(2.3 container-hours).
\end{itemize}
Finally, a \emph{prospective campaign} ran generated tests on 99 untested
helpful groups of plausible patches (10.8 container-hours), after we had
published its predictions. All controlled execution took 103 container-hours.

\subsection{What Was Planned in Advance}
\label{sec:status}

The fresh campaign was pre-registered. The predictions of the prospective
campaign were published before it ran, and every other campaign was specified
in writing before its own results existed. Some analyses were designed after
we had seen results, including the pre-flight check itself, the fix-copy check
and the decision basis. Appendix~\ref{app:status} lists the status of every
analysis.
\section{Results}
\label{sec:results}

\subsection{RQ1: How Often Are APR Comparisons Decided?}
\label{sec:rq1}

\paragraph{Defects4J} The labels already in the archive decide 12 of the 16
comparisons between the learned selector and its baselines. Only 3.9\% of the
candidates lack a label, yet the four comparisons the archive leaves open are
the most interesting ones, because they involve the selector's closest
competitors: its own refit without configuration identities, a
single-stage version, the testability product, and the top candidate of the best single configuration.

The usual convention hides this. Treating every unlabelled patch as incorrect gives a single number for each comparison, and Table~\ref{tab:convention} shows what that number would have been. The selector would have led the testability product by 1.02 points, the single-stage model by 1.84 and the source-agnostic refit by 3.11, and tied with the best configuration. A paper following the convention would therefore have reported three wins that the archive does not support, since each of the three intervals includes zero. The later evidence happened to confirm all three, but at the time nothing in the archive did.

\begin{table}[t]
\caption{The four Defects4J comparisons that the archive leaves open, and what the usual convention would have reported for them. ``Convention'' treats every unlabelled patch as incorrect; it equals the all-wrong value $\Delta_0$. Values are percentage points, selector minus baseline, on the archived labels.}
\label{tab:convention}
\small
\begin{tabular}{@{}lrll@{}}
\toprule
Comparison & Convention & Decision interval & Status \\
\midrule
Testability product & $+1.02$ & $[0.00,\,+1.43]$ & open \\
Single-stage model & $+1.84$ & $[0.00,\,+2.25]$ & open \\
Source-agnostic refit & $+3.11$ & $[-1.57,\,+4.34]$ & open \\
Best configuration & $0.00$ & $[-1.23,\,+2.25]$ & open \\
\bottomrule
\end{tabular}
\end{table}

\begin{figure}[t]
\centering
\begin{tikzpicture}[font=\footnotesize, x=0.5cm, y=0.42cm]
  \draw[-{Latex}] (0,0) -- (15.6,0) node[right] {pp};
  \foreach \x in {0,5,10,15} {
    \draw (\x,0.12) -- (\x,-0.12) node[below] {\ifnum\x>0 $+\x$\else $\x$\fi};
  }
  \draw[dashed, opengray] (0,0) -- (0,17.4);
  \foreach \i/\name/\lo/\hi/\when/\col in {
     1/Best configuration/1.23/1.64/fix re-run/badred,
     2/Testability product/0.41/0.61/review/badred,
     3/One-stage model/0.41/0.82/generated tests/badred,
     4/Source-agnostic refit/1.84/3.28/census/badred,
     5/First global position/5.94/6.35/archive/decblue,
     6/{Pooled order, then occurrence}/7.41/9.12/archive/decblue,
     7/Reciprocal-rank fusion/7.79/9.22/archive/decblue,
     8/Borda count/7.99/9.22/archive/decblue,
     9/Occurrence count/8.09/9.43/archive/decblue,
    10/Pooled order (MRA)/9.27/10.33/archive/decblue,
    11/Naturalness/10.86/12.91/archive/decblue,
    12/Entropy delta/11.27/12.70/archive/decblue,
    13/Token similarity/11.41/13.27/archive/decblue,
    14/CodeT5+ similarity/12.26/13.21/archive/decblue,
    15/Uniform random/12.47/14.71/archive/decblue,
    16/Mean normalised position/13.00/13.43/archive/decblue} {
    \node[anchor=east] at (-0.3,\i) {\name};
    \draw[line width=3.2pt, \col] (\lo,\i) -- (\hi,\i);
  }
  \draw[line width=3.2pt, decblue] (6.6,-2.0) -- (7.4,-2.0);
  \node[anchor=west] at (7.5,-2.0) {decided by the archive};
  \draw[line width=3.2pt, badred] (6.6,-2.8) -- (7.4,-2.8);
  \node[anchor=west] at (7.5,-2.8) {decided by added evidence};
\end{tikzpicture}
\caption{The 16 Defects4J comparisons after all evidence, as decision
intervals for the learned selector minus each baseline, in percentage points
of success rate. Blue comparisons are already decided by the archive. Red ones
were decided by added evidence, and their intervals sit much closer to zero.
Appendix~\ref{app:tables} gives every interval after every layer.}
\label{fig:d4j}
\end{figure}
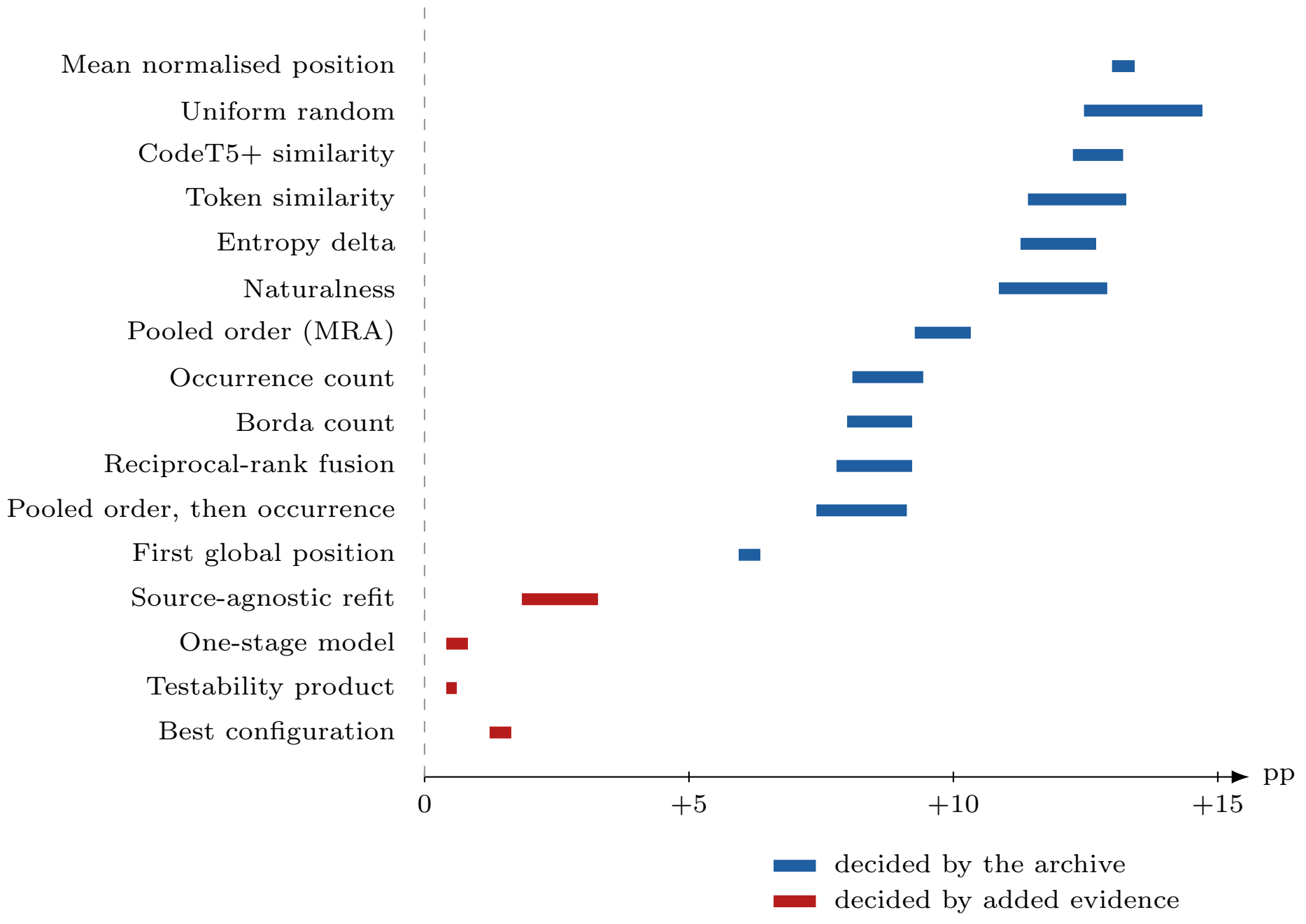

Each later evidence layer except the top-up decides
one more of them, and the final evidence decides all 16
(Table~\ref{tab:layers}). Figure~\ref{fig:d4j} shows the final intervals. The
selector wins every comparison, with lower bounds between $+0.41$ and $+13.00$
points.

\paragraph{How strong the decisions are} The decisions fall into two clear
groups. The twelve that the archive decides would need 29 to 64 wrong labels
to reopen, and eleven of them rest only on matches with the developer's fix
and execution failures, with no human judgment involved. The four decided
later are fragile: two wrong labels would reopen the testability and
single-stage comparisons, six the best-configuration comparison, and nine the
source-agnostic refit. A reader who distrusts the added evidence can still
rely on the twelve.

\paragraph{Do the decisions depend on our choices?} We checked this in several
ways, summarised here and detailed in Appendix~\ref{app:robustness}.
\begin{itemize}
\item Removing any single evidence layer reopens at most one comparison.
Without the census, the source-agnostic comparison reopens; without the
review, the testability comparison; without the fix re-run, the
best-configuration comparison. Removing the top-up or the generated tests
changes nothing.
\item The same 16 are decided when candidates are not pooled across prompts
and when copies that differ only in local names count as identical.
\item Four baselines often propose the unchanged buggy method, in 58 to 63\%
of bugs. If such candidates are made ineligible for both policies, the archive
decides 9 comparisons instead of 12, and the final evidence still decides all
16.
\item If we require the selector to win by at least one point instead of by
any margin, 14 of the 16 decisions hold. The twelve that the archive decides
still hold at five points.
\end{itemize}

\paragraph{HumanEval-Java} Here the archive's own rule for combining labels
matters a great deal. Under it, the comparison of the selector with the pooled
order ranges from $-16.18$ to $+29.22$ points, mostly because of 170
candidates that a reviewer had called correct while some archived run had
failed them. All 170 passed every test in both of our runs. With equivalence
to the developer's fix and our re-runs, the interval narrows to
$[+9.72, +21.23]$, and four of the six comparisons are decided, against one in
the archive. In one of them, the selector loses to its own first two stages.

\paragraph{The other settings} Elsewhere, few comparisons are decided
(Table~\ref{tab:settings_decided}). The most striking result is a reversal.
The selector that wins every comparison on Defects4J loses to plain
occurrence counting on RepairBench, $[-8.43, -2.25]$, and on D4C when success
means passing the bug's failure-revealing tests, $[-8.80, -0.54]$. Its
features describe where a candidate appears in the configurations it was
trained on, and that information does not carry over to other benchmarks.
Among plausible patches, the archived and re-run labels decide none of 21
comparisons, and all our evidence decides 14, several of them by a fraction
of a point. Appendix~\ref{app:other} covers the remaining settings, including
the published comparisons in RepairLLaMA's paper, of which 22 of 23 hold under
every possible value of the unknown labels.

\begin{table}[t]
\caption{Decided comparisons in each setting at the first and the last
evidence view, with one example comparison. ``vs $X$'' means the learned
selector minus $X$. LLM4PC and DL4PC are LLM4PatchCorrect and
DL4PatchCorrectness.}
\label{tab:settings_decided}
\small
\begin{tabularx}{\linewidth}{@{}lrlY@{}}
\toprule
Setting & Pairs & Decided & Example comparison, last interval (pp) \\
\midrule
Defects4J & 16 & 12 $\rightarrow$ 16 & vs pooled order, $[+9.27,\,+10.33]$ \\
HumanEval-Java & 6 & 1 $\rightarrow$ 4 & vs pooled order, $[+9.72,\,+21.23]$ \\
Fresh campaign & 5 & 0 $\rightarrow$ 0 & vs occurrence count, $[-3.53,\,+3.23]$ \\
RepairBench & 9 & 1 & vs occurrence count, $[-8.43,\,-2.25]$ \\
D4C & 4 & 1 & vs occurrence count, $[-8.80,\,-0.54]$ \\
Plausible patches & 21 & 0 $\rightarrow$ 14 & vs PrevaRank, $[+0.41,\,+7.58]$ \\
PrevaRank input order & 51 & 0 & hash vs supplied order, $[-13.44,\,+9.25]$ \\
Overfitting detectors & 10 & 0 & LLM4PC vs DL4PC, $[-11.45,\,+42.32]$ \\
\bottomrule
\end{tabularx}
\end{table}

\answerbox{RQ1}{The archive alone decides 12 of 16 Defects4J comparisons, and
added evidence decides the remaining four, but those four are fragile. Outside
Defects4J, few comparisons are decided, and the selector that wins on
Defects4J loses to occurrence counting on two other benchmarks.}

\subsection{RQ2: Can We Predict Which Comparisons Testing Will Settle?}
\label{sec:rq2}

\paragraph{Testing never moved the all-wrong value} None of the three test
campaigns on Defects4J refuted a group already known to be correct, so the
refutation limit applies to each of them. As the proposition predicts, the
all-wrong value of all 16 comparisons stayed exactly the same through the
census, the top-up and the generated tests. It changed only with the human
review and the fix re-run, the two layers that can show a patch is correct.

\paragraph{The four late comparisons} Table~\ref{tab:preflight} applies the
pre-flight check of Figure~\ref{fig:preflight} to each comparison that was
open when a campaign started. Each outcome explains why a comparison was or
was not decided.
\begin{itemize}
\item \textbf{Best configuration.} Its all-wrong value was zero, so no test
campaign could decide it, and none did. It was decided by the 0.8-hour fix
re-run, which withdrew false failures and turned eight of its locked groups
into correct ones.
\item \textbf{Testability product.} Its helpful groups were locked for
re-running. The generated tests could have decided it with any one of five
helpful groups, but none of the five was among their targets.
\item \textbf{Single-stage model.} The census and the top-up did not target
any of its three helpful groups. The generated tests did, and one failure, in
a candidate for bug Lang-45, decided it.
\item \textbf{Source-agnostic refit.} The census needed 8 helpful failures and
found 12.
\end{itemize}

\begin{table}[t]
\caption{The pre-flight check applied after the fact to the four Defects4J
comparisons that were open in the archive. Each cell gives the outcome for one
campaign, using the names of Figure~\ref{fig:preflight}. ``Locked'' groups
have conflicting archived results that a re-run cannot resolve. ``--'' means
already decided.}
\label{tab:preflight}
\small
\begin{tabularx}{\linewidth}{@{}>{\raggedright\arraybackslash}p{2.7cm}YYYY@{}}
\toprule
Comparison & Census (16.0 h) & Top-up (5.7 h) & Generated tests (13.0 h) & Fix re-run (0.8 h) \\
\midrule
Best configuration & cannot decide (all-wrong value is 0) & cannot decide (all-wrong value is 0) & cannot decide (all-wrong value is 0) & decided (confirming) \\
Testability product & cannot decide (helpful groups locked) & cannot decide (helpful groups locked) & targets miss it (0 of 5) & -- \\
One-stage model & targets miss it (0 of 3) & targets miss it (0 of 3) & decided (1 needed, 1 failed) & -- \\
Source-agnostic refit & decided (8 needed, 12 failed) & -- & -- & -- \\
\bottomrule
\end{tabularx}
\end{table}

\paragraph{Effort that could not help} The top-up spent 5.7 container-hours,
found 109 new failing patches, and decided nothing, because its targets could
reach none of the comparisons still open. Of the generated tests' 13.0 hours,
only 0.6 ran in jobs that touched a helpful group. Together, 18.1 of these
18.7 hours could not have decided an open comparison, and the check would have
shown this in advance. Across all three campaigns, 34.7 container-hours could
not decide the best-configuration comparison, which the 0.8-hour fix re-run
settled.

\paragraph{When the targets are right but too few fail} The fresh campaign
shows the last outcome of the check. Its main comparison, the selector against
occurrence counting, had an all-wrong value of $+0.05$ points, so testing
could decide it only by refuting at least 157 of its 159 helpful groups. Our
run refuted 133 and narrowed the interval from $[-17.41, +25.27]$ to
$[-3.53, +3.23]$, but at least 24 of the 26 remaining helpful groups would
have had to fail, and 18 of them had passed their failure-revealing tests.

\paragraph{Predicting outcomes in advance} The planner of
Section~\ref{sec:polarity} estimates the chance that a campaign decides a
comparison from the failure rate in a small pilot. Applied to our earlier
campaigns using a random fifth of each target list, it gave probability zero
to 13 comparisons, and all 13 stayed open. It gave 0.96 and 0.34 to the two
that were decided. We then tested it prospectively. Before running generated
tests on the 99 untested helpful groups of the plausible patches, we published
its predictions for the eight comparisons still open there.
Table~\ref{tab:prospective} shows that all eight came true. The one
comparison predicted to be decided, the selector against PrevaRank, was
decided by the first group that failed, and the seven others stayed open.
Only 7 of the 99 targeted groups failed, half the rate we had assumed.

\begin{table}[t]
\caption{The prospective test. Predictions were published before the
campaign ran. ``Predicted chance'' is the planner's probability that the
campaign decides the comparison. $^{a}$All-wrong value is zero, so testing cannot decide it.}
\label{tab:prospective}
\small
\begin{tabularx}{\linewidth}{@{}Yrll@{}}
\toprule
Comparison & Predicted chance & Interval after (pp) & Outcome \\
\midrule
Selector vs PrevaRank & 0.999 & $[+0.41,\,+7.58]$ & decided \\
PrevaRank vs pooled order & 0.098 & $[-5.61,\,+0.51]$ & open \\
APPT vs pooled order & 0.025 & $[-8.33,\,+1.10]$ & open \\
Selector vs CodeT5+ similarity & 0.004 & $[-1.48,\,+6.10]$ & open \\
Correctness stage vs CodeT5+ similarity & 0.003 & $[-1.59,\,+6.81]$ & open \\
Selector vs its correctness stage & 0$^{a}$ & $[-1.64,\,+1.02]$ & open \\
PrevaRank vs occurrence count & 0 & $[-5.66,\,+2.16]$ & open \\
APPT vs occurrence count & 0 & $[-8.03,\,+2.40]$ & open \\
\bottomrule
\end{tabularx}
\end{table}

\answerbox{RQ2}{Yes. Testing never moved the all-wrong value, as the
refutation limit predicts. The pre-flight check explains every comparison our
campaigns did or did not decide, and all eight predictions published before a
new campaign came true.}

\subsection{RQ3: How Trustworthy Are the Archived Test Results?}
\label{sec:rq3}

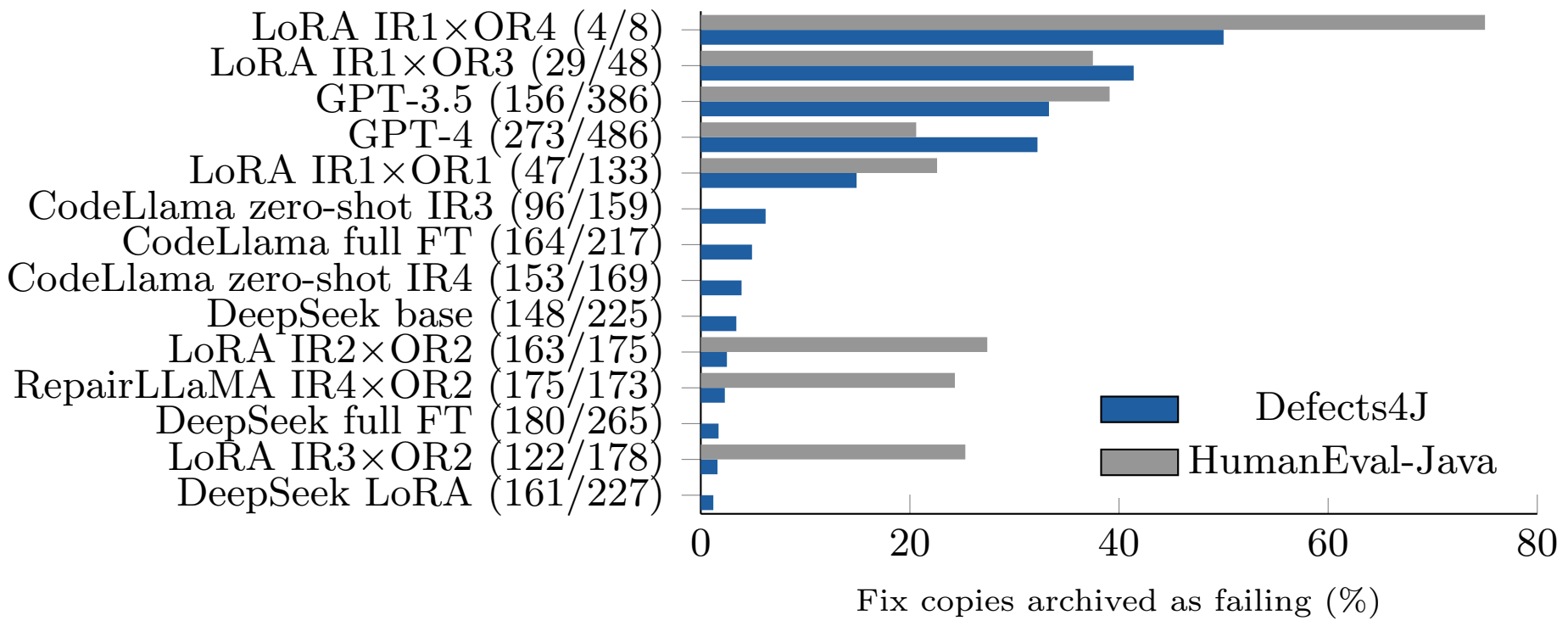
\begin{figure}[t]
\centering
\begin{tikzpicture}
\begin{axis}[xbar=0pt,bar width=3.3pt,width=0.62\linewidth,height=5.5cm,
  xmin=0,xmax=80,xlabel={Fix copies archived as failing (\%)},xlabel style={font=\footnotesize},
  ytick={0,...,13},yticklabels={{DeepSeek LoRA (161/227)},{LoRA IR3$\times$OR2 (122/178)},{DeepSeek full FT (180/265)},{RepairLLaMA IR4$\times$OR2 (175/173)},{LoRA IR2$\times$OR2 (163/175)},{DeepSeek base (148/225)},{CodeLlama zero-shot IR4 (153/169)},{CodeLlama full FT (164/217)},{CodeLlama zero-shot IR3 (96/159)},{LoRA IR1$\times$OR1 (47/133)},{GPT-4 (273/486)},{GPT-3.5 (156/386)},{LoRA IR1$\times$OR3 (29/48)},{LoRA IR1$\times$OR4 (4/8)}},
  y tick label style={font=\scriptsize},x tick label style={font=\scriptsize},
  enlarge y limits=0.04,area legend,legend style={font=\scriptsize,at={(0.98,0.03)},anchor=south east,draw=none},
  axis x line*=bottom,axis y line*=left]
\addplot[fill=decblue,draw=none] coordinates {(1.2,0) (1.6,1) (1.7,2) (2.3,3) (2.5,4) (3.4,5) (3.9,6) (4.9,7) (6.2,8) (14.9,9) (32.2,10) (33.3,11) (41.4,12) (50.0,13)};
\addplot[fill=opengray,draw=none] coordinates {(0.0,0) (25.3,1) (0.0,2) (24.3,3) (27.4,4) (0.0,5) (0.0,6) (0.0,7) (0.0,8) (22.6,9) (20.6,10) (39.1,11) (37.5,12) (75.0,13)};
\legend{Defects4J,HumanEval-Java}
\end{axis}
\end{tikzpicture}
\caption{Share of generated copies of the developer's fix that the RepairLLaMA
archive records as failing, per configuration (numbers of copies on Defects4J
and HumanEval-Java in parentheses; IR$k$ and OR$m$ are RepairLLaMA's input and
output formats). Defects4J is restricted to the 483 bugs whose fix lies inside
the replaced method.}
\label{fig:controls}
\end{figure}

\paragraph{Copies of the developer's fix} In the RepairLLaMA archive, 201 of
1,871 copies of the developer's fix (10.7\%) are recorded as failing on
Defects4J, and 440 of 2,849 (15.4\%) on HumanEval-Java
(Figure~\ref{fig:controls}). For comparison, the archive records none of its
585 GitBug-Java copies as failing, and RepairBench records 56 of 2,197
(2.5\%). On Defects4J the failures concentrate in a few configurations.
GPT-4 and GPT-3.5 lose about a third of their copies, mostly to recorded
compile failures (82 of GPT-4's 88), while the RepairLLaMA model loses 2.3\%.

The difference between models is not caused by which bugs they attempted. On
the 51 bugs where GPT-4 and the RepairLLaMA model both produced copies, GPT-4's
copies are recorded as failing in 33.5\% of cases and the RepairLLaMA model's
in 2.5\%. With each bug weighted equally, GPT-4's failure rate is 27.5 points
higher, with a 95\% bootstrap interval from 18.2 to 37.5 points. A broken
build of the whole bug does not explain it either: in 101 of the 107
bug-configuration pairs with a failing copy, the configuration's other outputs
for that bug compile.

\paragraph{The failures do not reproduce} We re-ran 80 archived failures of
fix copies. Of the 71 that reached a verdict, 68 passed the failure-revealing
tests in both runs. On HumanEval-Java, all 185 failing copies whose exact text
we ran passed every test.

\paragraph{Other failures of the same configurations are mostly real} The
failure audit re-ran a random sample of 160 other archived failures from the
five configurations whose fix copies fail more than 10\% of the time. Of these, 147 failed again,
5 passed, and 8 could not be resolved. Among resolved cases, 3.3\% did not
reproduce, with a 95\% interval of 1.1 to 7.5\%. We then simulated
withdrawing each failure label of these configurations at the upper end of
that interval. Each of the 16 decisions held in at least 95.9\% of 20,000
simulated draws, and the twelve robust ones in at least 99.99\%.

\paragraph{Other evidence} The census failures were reliable: re-run in the
original Defects4J 2.0 and Java 8 environment, 181 of 185 reproduced and none
passed. The random panel of 100 accepted-correct candidates produced no false
failure in 90 cases that could be evaluated. Only five of the 16 decisions
need any human judgment, and all 16 stay decided when every label that rests
on our own annotators' ``incorrect'' verdicts is set back to unknown.
Appendix~\ref{app:robustness} gives these checks in full.

\answerbox{RQ3}{Archived failures of fix copies are mostly harness errors, and
their rate differs sharply between models: 33.5\% for GPT-4 against 2.5\% for
the RepairLLaMA model on the same bugs. Other archived failures are mostly
real, and every decision survives the measured error rates.}
\section{Discussion}
\label{sec:discussion}

\paragraph{Report the interval, and what it rests on} A single success rate
hides whether the result depends on unchecked patches. The decision interval
answers that question directly, and the two strength measures tell a reader
how much to trust a decision. In our study they separate twelve decisions that
29 or more wrong labels would reopen from four that two to nine wrong labels
would reopen. They also point a reviewer to the 31 human-judged groups on
which five decisions depend, so a skeptical reader knows exactly which labels
to check.

\paragraph{Check before you test} Most evidence that can be added to an APR
archive cheaply is testing, and testing can only refute. Before starting a
campaign, a researcher can run the pre-flight check on its target list. The
check says which comparisons tests can never decide, which ones the planned
targets miss, and how likely the campaign is to decide the rest. In our study
it would have saved most of 18.7 container-hours, and its published
predictions all came true. Where testing cannot help, the useful evidence is
evidence that a patch is correct. In our case the cheapest such evidence was a
0.8-hour re-run of the developer's fix copies, which settled a comparison that
34.7 container-hours of testing could not.

\paragraph{Calibrate the harness with its own fix copies} Counting the copies
of the developer's fix that an archive records as failing takes minutes and
needs no execution. In the RepairLLaMA archive the rate ranges from 1.2\% to
more than 30\% across configurations on Defects4J, so the harness treats the
models being compared differently. We suggest that papers reporting per-model
results also report this rate. The same count shows which archived failures
deserve a re-run. Labels that record a skipped review should be released as
missing. In the detector benchmark, for example, 79 labels of overfitting
were assigned without review once a correct sibling patch had been found.

\paragraph{What the decided comparisons say about selectors} On Defects4J, a
selector trained on the archive's metadata beats all 16 baselines under every
possible value of the unknown labels, and eleven of these wins rest only on
matches with the developer's fix and on execution failures. On RepairBench,
and on D4C when success means passing the failure-revealing tests, the same
selector loses to plain occurrence counting. This echoes the finding of
\citet{durieux2019empirical} that Java repair tools fix 47\% of Defects4J bugs
but only 10 to 30\% of the bugs in other benchmarks. Studies that train and
evaluate a selector on one archive should report at least one transfer
setting. Among plausible patches, archived and re-run labels separate none of
the selectors, while generated tests, review and re-runs together decide 14
of 21 comparisons, so claims about ranking plausible patches need that kind of
evidence.

\paragraph{Beyond Java and beyond repair} The interval needs only two things: a label, or the absence of one, for each candidate, and each policy's choice on each bug. Any archive that releases per-candidate outcomes can be analysed this way, in any language. The refutation limit applies to every evaluation that counts a candidate as solved when no test rejects it, including agent leaderboards such as SWE-bench: such a score is the all-correct value of the interval, and stronger tests can only move it toward the all-wrong value \citep{yu2025utboost,yu2026sweabs}. The fix-copy check applies wherever a model can regenerate the reference fix and the archive records the test outcome of each candidate.

\needspace{12\baselineskip}
\paragraph{How to apply this in an evaluation} Five steps cover what this paper adds; the replication package contains the scripts behind the first four.
\begin{enumerate}
\item Group identical candidates, pool each bug's candidates, and compute the decision interval of every comparison from the labels you already have. Report each comparison as decided or open, with the interval.
\item For each decided comparison, report how many wrong labels would reopen it and which label sources its decision rests on.
\item Before any test campaign, run the pre-flight check on its target list. Drop the comparisons it cannot reach, and estimate from a small pilot how likely it is to decide the rest.
\item Count the copies of the developer's fix that the archive records as failing, per model. Report the rate, and re-run the failures in the cells where it is high.
\item Report at least one setting other than the one the technique was developed on.
\end{enumerate}
\section{Related Work}
\label{sec:related}

\paragraph{Patch correctness and its evidence} \citet{qi2015plausibility}
showed that most patches reported by early generate-and-validate repair tools
were incorrect, and that errors in the evaluation infrastructure had let many
of them pass as plausible. \citet{long2016analysis} found plausible patches far
outnumbering correct ones in the search spaces of such tools, and
\citet{smith2015cure} tied overfitting to the test suite used for repair.
\citet{le2019reliability} compared the common ways of labelling patches with a
gold set from independent developers. Author labels differed from that set on
13 of 118 patches, seven of which the authors had labelled unknown, and
independently generated tests exposed only a small share of the incorrect
patches. \citet{qin2021flaky} showed that flaky tests, most of them
deterministic within a fixed environment, change which bugs repair tools are
counted as fixing.

A large body of work predicts the missing labels. Patch correctness assessment
has used execution similarity \citep{xiong2018identifying}, embeddings of code
changes \citep{tian2020evaluating}, fine-tuned code models
\citep{zhang2024appt,zhang2026compass}, invariants \citep{lecong2023invalidator}
and language models given context \citep{zhou2024llm4patchcorrect}. The
measured accuracy of these predictors changes with the projects, tools and
patch mix of the evaluation set
\citep{wang2020automated,yang2023review,williams2026pod}. Historian keeps an
explicit Unknown verdict when historical evidence is insufficient
\citep{moslemi2026historian}. All of this work tries to make the labels better. None of it asks whether a reported ranking depends on the patches that were never labelled, and we know of no APR evaluation that reports that dependence. We take the labels as they are and ask which comparisons they decide.

\paragraph{Evidence that refutes} Several techniques generate tests to show
that a patch is wrong. DiffTGen searches for inputs on which a patch and the
developer's fix behave differently \citep{xin2017identifying},
\citet{ye2021automated} assess patches with tests generated on the fixed
program, and FixCheck adds assertions written by a language model to random
tests \citep{molina2024fixcheck}. For coding agents, PatchDiff finds
behavioural differences between accepted SWE-bench patches
\citep{jimenez2024swebench} and the reference, only part of which its manual
check confirmed as defects \citep{wang2025patchdiff}. UTBoost and SWE-ABS
strengthen the benchmark's tests and change the leaderboard ranking
\citep{yu2025utboost,yu2026sweabs}, and \citet{xu2026validation} replay agents'
validation commands on the buggy, candidate and reference states, much as our
re-runs use fixed and buggy controls. Such tests can show a patch wrong, but
passing them does not show it correct. A leaderboard that counts every
unrefuted patch as resolved therefore reports each score's all-correct value,
and further valid refutations, with accepted labels fixed, can only move it
toward the all-wrong value.

\paragraph{Incomplete judgments and partial identification} Information
retrieval faced the same problem earlier. Pooling leaves many relevant
documents unjudged, although system rankings proved stable
\citep{zobel1998reliable}, and measures such as bpref, infAP and the residual of
rank-biased precision respond to it
\citep{buckley2004bpref,yilmaz2006infap,moffat2008rbp}. Minimal test
collections state a worst-case condition under which the sign of a paired
difference holds whatever the unjudged documents are, although their
experiments stop on a probabilistic version of that condition
\citep{carterette2006mtc}. Targeted judging compares systems through residual
bounds of rank-biased precision \citep{moffat2007strategic}. Our decision
interval applies the worst-case condition to success@1, with groups of
identical patches, bug-level pools and random tie-breaking in place of
documents and rankings. What changes in APR is the evidence, which mostly
refutes.

Bounds that make no assumption about missing outcomes go back to
\citet{manski2003partial} and are surveyed by \citet{tamer2010partial};
\citet{imbens2004confidence} and \citet{stoye2009more} give confidence
intervals for them. \citet{kasy2016partial} characterises when the ranking of
two policies is identified even though their effects are not. Work on
selective labels evaluates policies, and bounds their regret or predictive
performance, when only some outcomes are observed
\citep{lakkaraju2017selective,kallus2018confounding,rambachan2026robust}. We
take the assumption-free end of this family, because the labels in an APR
archive are those that an earlier pipeline chose to check. Label errors change
model rankings on machine-learning benchmarks \citep{northcutt2021pervasive},
and fragility indices count how few changes overturn a conclusion
\citep{walsh2014fragility,giordano2026automatic}, as our count of the label
flips that reopen a decision does for known labels.

\paragraph{Validity of leaderboard comparisons} \citet{huang2026leaderboard}
finds that close ranks on agent leaderboards are often unresolved once the
estimand, the label source and the uncertainty rule are stated.
\citet{liu2026converged} find that exact paired tests separate none of the 29
adjacent pairs among the top 30 SWE-bench Verified submissions. Closest to our
setting, \citet{kuboyama2026labels} uses the criterion we adopt, a winner
under every possible value of the unknown labels, and bounds the number of
labels needed to fix the winner among selective predictors. There, a label
that is read can turn out either way. In APR, a test run returns ``incorrect''
or nothing, and we show (Proposition~\ref{prop:limit}) that when a
comparison's all-wrong value is zero, no number of such labels is enough.

\paragraph{Selecting among generated programs} Selection methods for code
generation use rankers trained on execution \citep{inala2022coderanker,lyu2025toppass},
agreement among samples \citep{wang2023selfconsistency} and with generated
tests \citep{chen2023codet}, inputs that distinguish candidates
\citep{li2025sstar}, and naturalness \citep{xia2023automated,yang2024naturalness}.
For repair, Prophet ranks candidates before validation with a learned model of
correct code \citep{long2016prophet}, PrevaRank ranks plausible patches by
historical fix features \citep{bhuiyan2026prevarank}, and imperfect verifiers
cap the benefit of resampling \citep{stroebl2026limits}.
\section{Threats to Validity}
\label{sec:threats}

\paragraph{Construct validity} Our main measure is success@1 with ties broken
uniformly at random. The bounds for success within three and five attempts
agree with it against the pooled order, but not against occurrence count or
Borda count. Grouping assumes that candidates with the same syntax tree,
placed in the same method of the same program, behave the same, which a test
that observes line numbers could break. The 16 decisions do not change when
candidates are not pooled across a bug's prompts, or when copies that differ
only in local names count as identical.

A reference-behaviour witness shows that a candidate behaves differently from
the developer's fix on a generated input. Whether that difference violates the
specification is the test-oracle problem \citep{barr2015oracle}, and no final
decision needs a witness. Human verdicts compare a candidate with the
developer's fix. The archive's reviewers checked semantic equivalence to it,
and our annotators' strict rule counted a different but valid repair as
incorrect. Their verdicts on eight groups enter four decision bases, but only
one decision depends on them, and none needs their ``incorrect'' verdicts.
Both equivalence and the strict rule credit the developer's fix itself, which
the models may have memorised, so a decision can partly reflect which policy
finds a memorised fix first. The fix-copy check covers copies of the fix only
and does not show why a copy failed.

Three published detectors could not serve as baselines. ComPass has an expired
artifact and training data that contain Defects4J fixes
\citep{zhang2026compass}, the retriever weights of LLM4PatchCorrect are no
longer available \citep{zhou2024llm4patchcorrect}, and Invalidator needs the
developer's fix and about seven minutes per patch \citep{lecong2023invalidator}.

\paragraph{Internal validity} The fresh campaign and the prospective campaign
are confirmatory (Appendix~\ref{app:status}). An optional further stage of the
fresh campaign was not completed, and we report its pre-registered G-E1
result. The pre-flight check was designed after the earlier campaigns ran, and
we apply it to their frozen target lists. Evidence layers were added one
after another, each fixed before its outcome but without a preset stopping
point, and the fix re-run was specified after a view that counted
fix-equivalent groups as correct had shown which comparison they could decide.

The human review was carried out by two annotators, the author and a second
annotator, whose labels agreed on all 104 items. The only decision that
depends on their verdicts, against the testability product, rests on two
groups that were both judged correct. Our re-runs used Defects4J 3.0.1 and Java 11
instead of the archive's original environment, and flaky tests can pass under
one Java version and fail under another \citep{qin2021flaky}. The
cross-version re-run and the false-failure panel bound this effect. The
selector learns from archived outcomes and configuration identities, so it
could profit from configuration-specific false failures; the failure audit
limits this possibility (Section~\ref{sec:rq3}).

\paragraph{External validity} All settings are in Java, and all but the
fresh campaign reuse outputs released by others. The language models behind
these outputs may have seen Defects4J during training
\citep{ramos2025memorizing}. We chose the settings because their artifacts
were available, so we make no claim about APR in general.

\paragraph{Conclusion validity} A decision interval describes one fixed set
of candidates and contains no sampling uncertainty. To estimate how much the
results would vary with different bugs, we refit all three stages of the
selector in each of 2,000 bug-level bootstrap replicates for the comparison
with the pooled order after the census. This gives a nominal 95\%
Imbens--Manski interval \citep{imbens2004confidence} of $[+3.38, +14.06]$
points, with a positive lower bound in 99.95\% of refits, and leaving out one
bug at a time keeps the lower bound between 4.87 and 6.03. In our coverage
simulation (five scenarios, 1,000 replicates each, 500 bootstrap draws per
replicate), the worst-endpoint coverage was 0.931 to 0.935 in four scenarios
against a nominal 0.95. As our protocol prescribed, every claim therefore rests
on the decision intervals and their flips, and the bootstrap only supports
them.
\section{Conclusion}
\label{sec:conclusion}

Missing correctness labels do not have to leave an APR comparison unresolved.
By computing the full range of a comparison over every possible value of the
unknown labels, we can report it as decided or open, and say how many wrong
labels it would take to change that. On RepairLLaMA's Defects4J archive, this
decided all 16 comparisons of a learned selector with its baselines, twelve
robustly and four narrowly, and all 16 hold without our annotators'
``incorrect'' verdicts.

Because testing can only show that a patch is wrong, the comparisons that no
test campaign can settle can be identified before the campaign starts, and the
campaign's target list shows which others it misses. In a prospective test,
all eight of the check's published predictions held. Copies of the
developer's fix, which must pass, reveal that archives reject correct patches
at very different rates for different models: 33.5\% of \mbox{GPT-4's} copies
against 2.5\% of the RepairLLaMA model's on the same bugs, while 97\% of the
other resolved failures of the affected configurations reproduce.

\backmatter

\section*{Declarations}
\pdfbookmark[1]{Declarations}{declarations}

\begin{itemize}
\item \textbf{Funding.} No funding was received for this work.

\item \textbf{Competing interests.} The author has no competing interests to
declare.

\item \textbf{Ethics approval and consent to participate.} Not applicable. The
study analyses publicly released software artifacts. The human review consisted
of two annotators labelling generated patches, and no data about people were
collected.

\item \textbf{Consent for publication.} Not applicable.

\item \textbf{Data and code availability.} The replication package is available
at \url{https://github.com/SNaf1/decided-or-open} and archived on Zenodo \citep{nafis2026replication}. It
contains the evidence ledgers with their provenance, the frozen protocols with
their hashes, the execution records, the scripts behind every number, and a
script that regenerates the tables and figures from the stored results and
checks the principal numbers. Defects4J 3.0.1 (commit \texttt{8c16da82},
Java 11), Defects4J 2.0.0 (commit \texttt{01f13c69}, Java 8) and HumanEval-Java
(JUnit 4.12 on Temurin 8) ran in containers pinned by digest. Upstream
artifacts that may not be redistributed are referenced by revision and hash,
with retrieval scripts that verify them.

\item \textbf{Use of AI tools.} An AI assistant helped write and run the
analysis scripts and screened the counterexamples of the generated-test
campaigns against the four exclusions fixed in advance
(Section~\ref{sec:layers}). Every verdict and its reason is recorded in the
replication package.

\item \textbf{Author contribution.} Sadnan Nafis designed the study, carried
out the experiments and analyses, and wrote the manuscript.
\end{itemize}

\begin{appendices}

\section{Full Procedures}
\label{app:procedures}

\paragraph{Controlled re-execution} The census, the top-up and the fix re-run
share one protocol. A pinned, offline container with Defects4J 3.0.1 and
Java 11 places each candidate by replacing the archived buggy method verbatim.
A missing or ambiguous placement leaves the candidate unresolved. The protocol
runs the fixed version, the buggy version and every candidate on the bug's
failure-revealing tests twice, the second time in reverse order. A test is
admissible when it is not on Defects4J's lists of flaky tests
\citep{just2014defects4j}, the fixed version passes it twice, and the buggy
version fails it twice with the same failure signature. A \emph{rejection
witness} fails an admissible test with identical signatures in both runs, with
no admissible test disagreeing between the two runs. Loading, linkage and
memory errors void the evidence, and a candidate that passes stays unknown. In
the census and the top-up, a candidate that twice fails to compile also stays
unknown, because the archive records these candidates as compiling.

\paragraph{Generated tests} The 165 groups are those whose members all passed
the archived tests and that bear on a comparison other than the one with
uniform choice. EvoSuite \citep{fraser2011evosuite}, with two seeds, and
Randoop \citep{pacheco2007feedback} generate tests on the fixed version of the
modified classes, in a Defects4J 2.0 and Java 8 container. Tests that fail on
the fix are removed, and one bug yielded no test suite. A test that passes on
the fix and fails on the candidate in both of two runs is a counterexample.
A counterexample is a \emph{reference-behaviour witness} unless the difference
is limited to one of the four cases listed in Section~\ref{sec:layers}. Each
counterexample was screened against these cases, and a witness marks its whole
group incorrect.

\paragraph{Human review} The 44 groups bore on one of six comparisons: the
four open in the archive, the pooled order and occurrence count. The 60 audit
items were randomly sampled archived judgments of test-passing candidates, 30
accepted as correct and 30 as incorrect. The packets did not show which
comparison a candidate bears on, which items were audit items, or their
archived labels; the author, who prepared the packets, had access to the audit
key. The verdicts apply to whole groups.

\paragraph{Fix re-run} The rename-aware fingerprint strips annotations and
renames parameters and local variables in order of declaration. Equivalence to
the developer's fix never overrides an execution result on its own.

\paragraph{HumanEval-Java} We executed all 2,230 unknown candidates relevant
to any comparison of the selector, with fixed and buggy controls and two
repetitions. The final view labels a candidate correct when it is equivalent
to the developer's fix (the same syntax tree, or a consistent renaming of
locals) or when it passes every test and every archived review calls it
correct. It labels a candidate incorrect when it fails a test or twice fails to
compile. For HumanEval-Java we keep the candidate-level rules under which the
archived comparisons were first computed, so that archived and re-run
intervals describe the same quantity.

\paragraph{Fresh campaign} Evidence starts from nothing. Matches with the
developer's fix are correct and unchanged code is incorrect (view G-E0). One
representative of every unknown group relevant to a pre-registered comparison
is then run under the census protocol, and a rejection witness or a twice-failed
compile counts as incorrect (view G-E1).

\paragraph{D4C} D4C releases no correctness labels, so its measure of success
is passing the bug's failure-revealing tests in our controlled runs of its
2,594 candidates.

\needspace{0.4\textheight}
\section{Status of Every Analysis}
\label{app:status}

Table~\ref{tab:status} gives the status of every analysis, from the pre-registered fresh
campaign to the analyses we designed after seeing results.

\begin{table}[!htb]
\caption{Status of each analysis.}
\label{tab:status}
\small
\begin{tabularx}{\linewidth}{@{}>{\raggedright\arraybackslash}p{3.4cm}Y@{}}
\toprule
Status & Analyses \\
\midrule
Pre-registered on unseen data & Fresh campaign: generation, policies, the main comparison (selector vs occurrence count), four secondary comparisons, views G-E0 and G-E1 \\
Predictions published before execution & Prospective generated-test campaign on plausible patches \\
Fixed before its outcome, on data already examined & Learned selector (trained before this study); application rules; census; code-reading baselines; plausible-patch setting; PrevaRank input orders; top-up; generated tests; cross-version re-run; false-failure panel; human review and audit; HumanEval-Java re-execution; rename-aware identity; fix re-run; D4C and RepairBench transfer; detector bounds; audit of published comparisons; failure audit \\
Designed after the results & Unchanged-code rule; full specification grid; refutation-limit analysis, pre-flight check and planner; fix-copy check and its bootstrap; decision basis; rubric sensitivity of the human verdicts; suspect-cell and stress views; error budget; unchanged-code-ineligible intervals; plausible-patch setting with the final evidence; leave-one-layer-out analysis \\
\bottomrule
\end{tabularx}
\end{table}

\needspace{0.4\textheight}
\section{Full Defects4J Results}
\label{app:tables}

Table~\ref{tab:d4j-full} gives every Defects4J comparison after every evidence
layer, with the strength measures of Section~\ref{sec:strength}. ``Flips'' is
the smallest number of wrong known labels that would reopen the final
decision, ``Human'' the number of human-judged groups in its decision basis,
and ``Break-even'' the label error rate at which its expected margin reaches
zero.

\begin{table}[!htb]
\caption{Decisions on Defects4J, learned selector minus each baseline (pp of
success rate, equal bug weights, 488 bugs). $\bullet$ decided, $\circ$ open,
after each layer of Table~\ref{tab:layers}.}
\label{tab:d4j-full}
\fontsize{7}{8.5}\selectfont\setlength{\tabcolsep}{2pt}
\begin{tabular}{@{}lccccccrrrr@{}}
\toprule
 & \multicolumn{6}{c}{Decided after evidence layer} & \multicolumn{4}{c}{Final evidence} \\
\cmidrule(lr){2-7}\cmidrule(l){8-11}
Baseline & E0 & E1 & E2 & E3 & E4 & E5 & Interval (pp) & Flips & Human & Break-even \\
\midrule
Pooled order & $\bullet$ & $\bullet$ & $\bullet$ & $\bullet$ & $\bullet$ & $\bullet$ & $[+9.27,\,+10.33]$ & 51 & 0 & 9.5\% \\
First global position & $\bullet$ & $\bullet$ & $\bullet$ & $\bullet$ & $\bullet$ & $\bullet$ & $[+5.94,\,+6.35]$ & 29 & 6 & 6.1\% \\
Borda count & $\bullet$ & $\bullet$ & $\bullet$ & $\bullet$ & $\bullet$ & $\bullet$ & $[+7.99,\,+9.22]$ & 39 & 0 & 9.6\% \\
Rank fusion & $\bullet$ & $\bullet$ & $\bullet$ & $\bullet$ & $\bullet$ & $\bullet$ & $[+7.79,\,+9.22]$ & 38 & 0 & 9.1\% \\
Mean position & $\bullet$ & $\bullet$ & $\bullet$ & $\bullet$ & $\bullet$ & $\bullet$ & $[+13.00,\,+13.43]$ & 64 & 0 & 12.0\% \\
Occurrence count & $\bullet$ & $\bullet$ & $\bullet$ & $\bullet$ & $\bullet$ & $\bullet$ & $[+8.09,\,+9.43]$ & 40 & 0 & 9.4\% \\
Pooled, then occurrence & $\bullet$ & $\bullet$ & $\bullet$ & $\bullet$ & $\bullet$ & $\bullet$ & $[+7.41,\,+9.12]$ & 37 & 0 & 8.8\% \\
Token similarity & $\bullet$ & $\bullet$ & $\bullet$ & $\bullet$ & $\bullet$ & $\bullet$ & $[+11.41,\,+13.27]$ & 56 & 0 & 10.8\% \\
CodeT5+ similarity & $\bullet$ & $\bullet$ & $\bullet$ & $\bullet$ & $\bullet$ & $\bullet$ & $[+12.26,\,+13.21]$ & 60 & 0 & 11.7\% \\
Naturalness & $\bullet$ & $\bullet$ & $\bullet$ & $\bullet$ & $\bullet$ & $\bullet$ & $[+10.86,\,+12.91]$ & 53 & 0 & 10.3\% \\
Entropy delta & $\bullet$ & $\bullet$ & $\bullet$ & $\bullet$ & $\bullet$ & $\bullet$ & $[+11.27,\,+12.70]$ & 55 & 0 & 10.5\% \\
Uniform random & $\bullet$ & $\bullet$ & $\bullet$ & $\bullet$ & $\bullet$ & $\bullet$ & $[+12.47,\,+14.71]$ & 62 & 0 & 11.3\% \\
\midrule
Source-agnostic refit & $\circ$ & $\bullet$ & $\bullet$ & $\bullet$ & $\bullet$ & $\bullet$ & $[+1.84,\,+3.28]$ & 9 & 9 & 3.5\% \\
One-stage logistic & $\circ$ & $\circ$ & $\circ$ & $\bullet$ & $\bullet$ & $\bullet$ & $[+0.41,\,+0.82]$ & 2 & 4 & 2.6\% \\
Testability product & $\circ$ & $\circ$ & $\circ$ & $\circ$ & $\bullet$ & $\bullet$ & $[+0.41,\,+0.61]$ & 2 & 6 & 2.0\% \\
Best configuration & $\circ$ & $\circ$ & $\circ$ & $\circ$ & $\circ$ & $\bullet$ & $[+1.23,\,+1.64]$ & 6 & 12 & 2.2\% \\
\midrule
Decided & 12 & 13 & 13 & 14 & 15 & 16 & 16 of 16 & & & \\
\bottomrule
\end{tabular}
\end{table}

\section{Robustness Details}
\label{app:robustness}

\paragraph{Unchanged-code candidates} Occurrence count, Borda count,
reciprocal-rank fusion and the pooled order with occurrence tie-breaking
propose the unchanged buggy method in 58 to 63\% of bugs. If such candidates
are ineligible for both policies, the lower bounds against these four
baselines fall to between $+1.13$ and $+2.66$ points, and those against the
other twelve move by at most 2.1 points. No decision rests on a label set by
the unchanged-code rule.

\paragraph{Specification grid} Across the grid of unit, identity, group-label
rule, unchanged-code treatment and weighting, 83 of the 90 cells for the
pooled order are decided after the census, including per-prompt units with
exact-text identity ($[+2.85, +8.16]$). The seven open cells all combine the
strict group-label rule, under which any unknown member leaves its group
unknown, with no unchanged-code rule. In a relaxation where unknown
test-passing groups are correct at their project's reviewed rate within
$\pm\delta$, the pooled-order comparison stays decided for every $\delta$,
while the best-configuration comparison stays open even at $\delta=0$.

\paragraph{Beyond the first attempt} Validation effort is often counted in
candidates validated \citep{liu2020efficiency}, so we also computed exact bounds
for success within three and five attempts and for the expected number of
attempts to the first correct patch, capped at five. Against the pooled order,
success within three and five attempts is open in the archive and decided
after the census ($[+2.36, +9.82]$ and $[+0.81, +8.47]$), and the selector
needs 0.15 to 0.39 fewer attempts per bug. These bounds are exact: within a
bug, each quantity depends on the labels only through the number of correct
groups in each tie block among the first five positions and is monotone in
those counts. Within five attempts, its comparisons with occurrence count
($[-3.88, +5.45]$), Borda count, the pooled order with occurrence tie-breaking
and its three ablations are open.

\paragraph{Suspect failure labels} Setting back to unknown the 533 archived
failure labels in bug-configuration pairs where a fix copy failed leaves all
16 decisions in place. Treating as uninformative every failure label of the
five configurations whose copies fail more than 10\% of the time (15,688
labels, a third of all failures) keeps 9 decided, three of the four late ones
among them. At the error rates measured for every label source, 14 decisions
hold in at least 99.7\% of simulated draws, and the testability and
single-stage decisions in 91\% and 93\%.

\paragraph{Rejection witnesses} The census witnesses reproduced in 181 of 185
cases in the original environment (97.8\%, Clopper--Pearson 95\% interval 94.6
to 99.4\% \citep{clopper1934use}). The other four became unresolved. Without
any census witness, only the source-agnostic comparison would reopen. The
random panel of accepted-correct candidates gave no false failure in 90
evaluable cases (95\% upper limit 4.0\%). The 612 reference-behaviour
witnesses from generated tests fall in 23 candidates of 11 bugs, 570 of them in
four Lang-45 candidates, and no final decision needs their labels.

\paragraph{Human judgments} Under their strict rule, our annotators judged 13
of the 30 accepted-correct audit candidates incorrect and 3 of the 30
accepted-incorrect ones correct. In several of the 13, the candidate differed
from the fix only in an exception message or in the order of floating-point
evaluation. In the decision bases, five of the 16 Defects4J decisions need any
human judgment, together 31 distinct groups in 27 bugs: 23 carry the archive's
own reviews and 8 our annotators' verdicts. Setting back to unknown every
candidate whose incorrect label rests on our annotators' verdicts alone (47
candidates in 21 groups) leaves all 16 decided, and the largest drop in a lower
bound is about one point, from $+5.94$ to $+4.92$ for the first global
position. Flipping all 16 labels that the audit disputed lowers none of the
lower bounds. Of the groups the annotators judged, none of the 18 judged
correct that received generated tests had a witness, against 4 of the 25
judged incorrect.

\section{Details of the Other Settings}
\label{app:other}

\paragraph{HumanEval-Java} The archive decides the comparison with the pooled
order only if reviewers' verdicts override recorded failures
($[+9.89, +21.19]$); otherwise it is open ($[-3.09, +6.63]$). Equivalence to
the developer's fix alone decides it ($[+0.18, +24.55]$). In the decided
comparison with the testability product, the selector loses,
$[-1.85, -1.23]$.

\paragraph{Plausible patches} On archived and census labels none of the 21
comparisons is decided, including those with PrevaRank ($[-4.41, +9.84]$) and
APPT ($[-4.10, +12.01]$). Before the prospective campaign, 13 of the 21 are
decided, among them the selector against the pooled order
($[+0.03, +2.87]$), occurrence counting and APPT, and its correctness stage
against PrevaRank ($[+0.10, +8.09]$). The prospective campaign adds the
selector against PrevaRank.

\paragraph{Transfer} D4C shares 246 of its 276 bugs with the Defects4J
training pool, and the fresh campaign uses Defects4J bugs, an overlap that may
favour the selector; RepairBench, where it also loses, shares none. Without
configuration identities, the selector picks the same candidate as the pooled
order in 267 of D4C's 276 bugs. In the fresh campaign, each secondary
comparison needed at least 1,176 refutations and received at most 1,109.

\paragraph{Published comparisons} Of the 23 pairwise directions between
configurations stated in RepairLLaMA's paper \citep{silva2025repairllama}, 22
hold under every possible value of the archive's unknown labels, and one
(GPT-4 over RepairLLaMA on HumanEval-Java) is open. Five hold on matches with
the developer's fix and execution failures alone, and 17 more only through
human judgments. Correcting for failing fix copies reverses none, reorders
three pairs that the paper does not discuss, and shrinks GPT-4's
HumanEval-Java lead from 7 bugs to 3.

\paragraph{Overfitting detectors} The benchmark of \citet{williams2026pod}
scores five detectors on 169 RepairLLaMA patches: entropy delta
\citep{yang2024naturalness}, Invalidator \citep{lecong2023invalidator},
FixCheck \citep{molina2024fixcheck}, LLM4PatchCorrect
\citep{zhou2024llm4patchcorrect} and DL4PatchCorrectness
\citep{tian2020evaluating}. Of its overfitting labels, 79 were assigned without
review once a correct sibling patch had been found. Treating these 79 as
unknown, with balanced accuracy bounded for each possible number of correct
patches among them, leaves all ten pairwise differences open; the largest,
10.23 points, spans $[-11.45, +42.32]$. The study's main finding holds:
random sampling matches or beats every detector on at least 85.7\% of the bugs
with a correct patch, under every possible value of the unknown labels.

\paragraph{Supplied order} PrevaRank's released tool keeps the supplied order
inside 74 to 79\% of its tie groups. Supplying candidates in the generator's
order gives a success-rate lower bound of 56.6\% against 45.9 to 53.5\% for
50 random orders, but none of the 51 paired comparisons is decided, so the
effect of the order is not identified.

\end{appendices}

\pdfbookmark[1]{References}{references}
\bibliography{references}

\end{document}